\documentclass[preprint,1p,a4paper,11pt]{elsarticle}
\biboptions{sort&compress}
\usepackage{amsmath,amssymb,mathtools}
\usepackage{graphicx}
\usepackage{booktabs,cellspace}
\usepackage{color}
\usepackage{hyperref}
\usepackage{xspace}
\usepackage{adjustbox}
\usepackage[normalem]{ulem}
\usepackage{geometry}
\usepackage{fancyhdr}       

\newcommand{\as}{\ensuremath{\alpha_s}\xspace}

\newcommand{\GeV}{\ensuremath{\,\mathrm{GeV}}\xspace}

\newcommand{\muR}{\ensuremath{{\mu_{\mathrm{R}}}}\xspace}
\newcommand{\muF}{\ensuremath{{\mu_{\mathrm{F}}}}\xspace}

\newcommand{\dd}{\mathrm{d}}
\newcommand{\NC}{\mathrm{NC}}
\newcommand{\CC}{\mathrm{CC}}
\newcommand{\hoppet}{{\sc hoppet}}
\newcommand{\apfel}{{\sc apfel}{\footnotesize ++}}

\newcommand{\order}[1]{\mathcal{O}\left(#1\right)}

\newcommand{\ydis}{y}
\newcommand{\xdis}{x_{\rm B}}

\definecolor{darkgreen}{rgb}{0,0.4,0}
\definecolor{grey}{rgb}{0.5,0.5,0.5}
\definecolor{orange}{rgb}{0.9,0.5,0.0}
\definecolor{lightblue}{rgb}{0.0,0.5,1.0}
\usepackage[dvipsnames]{xcolor}

\newcommand{\ttt}[1]{\texttt{#1}}
\newcommand{\repolink}[2]{\href{https://github.com/alexanderkarlberg/n3lo-structure-function-benchmarks/blob/main/#1}{\ttt{#2}}}

\newcommand{\auxlink}[1]{\repolink{aux/#1}{#1}}
\newcommand{\codelink}[1]{\repolink{code/#1}{#1}}

\newcommand{\nn}{\nonumber}
\newcommand{\hspn}{{\hspace{-3mm}}}
\def\z#1{{\zeta_{\:\! #1}}}
\def\zss{{\zeta_{2}^{\,2}}}

\def\zs{{\zeta_{2}^{\,2}}}
\def\ca{{C^{}_A}}
\def\cas{{C^{\, 2}_A}}

\def\cf{{C^{}_F}}
\def\cfs{{C^{\, 2}_F}}
\def\cft{{C^{\, 3}_F}}
\def\nf{{n^{}_{\! f}}}
\def\n2f{{n^{\,2}_{\! f}}}
\def\nfs{{n^{\,2}_{\! f}}}

\def\dabc2{{d^{\:\!abc}d_{abc}}}

\def\fl11{fl_{11}}
\def\fl02{fl_{02}}

\journal{European Physical Journal C}

\begin{document}
\allowdisplaybreaks
\begin{frontmatter}
\begin{flushright}
CERN-TH-2024-018
\end{flushright}
\title{Benchmark of \\ deep-inelastic-scattering structure functions at $\order{\as^3}$}

\author[1]{Valerio Bertone}\ead{valerio.bertone@cea.fr} 
\author[2]{Alexander Karlberg}\ead{alexander.karlberg@cern.ch}

\affiliation[1]{organization={IRFU, CEA, Paris-Saclay}, postcode={F-91191} ,city={Gif-sur-Yvette}, country={France}}
\affiliation[2]{organization={CERN, Theoretical Physics Department}, postcode={CH-1211} ,city={Geneva 23}, country={Switzerland}}

\begin{abstract}
  We present a benchmark comparison of the massless inclusive
  deep-inelastic-scattering (DIS) structure functions up to
  $\mathcal{O}(\as^3)$ in perturbative QCD. The comparison is
  performed using the codes \apfel{} and \hoppet{} within the
  framework of the variable-flavour-number scheme and over a broad
  kinematic range relevant to the extraction of parton distribution
  functions. We provide results for both the single structure
  functions and the reduced cross sections in both neutral- and
  charged-current DIS. Look-up tables for future reference are
  included, and we also release the code used for the benchmark.
\end{abstract}


\end{frontmatter}

\tableofcontents

\section{Introduction}\label{sec:introduction}

Deep inelastic scattering (DIS) is one of the best theoretically
understood processes in perturbative QCD, \textit{cf.}
Refs.~\cite{Ellis:1996mzs, Blumlein:2012bf} for a review.
Most significantly, it offers unique insight into the proton
structure, and to this day legacy DIS data still has a major impact on
fits of parton distribution functions
(PDFs)~\cite{NNPDF:2017mvq,NNPDF:2021njg, Bailey:2020ooq, Hou:2019efy,
  H1:2015ubc, Alekhin:2017kpj, ATLAS:2021vod}.

The inclusive DIS cross sections can be parametrised in terms of
structure functions.
These structure functions are inherently non-perturbative quantities,
but they can be expressed as convolutions between hard perturbative
coefficient functions and PDFs, where the latter encompass the
non-perturbative contribution. As of today, the massless DIS
coefficient functions are fully known up to $\mathcal{O}(\as^3)$,
allowing us to compute structure functions up to
next-to-next-to-next-to-leading order (N$^3$LO)
accuracy~\cite{SanchezGuillen:1990iq,vanNeerven:1991nn,Zijlstra:1992qd,Zijlstra:1992kj,vanNeerven:1999ca,vanNeerven:2000uj,Moch:1999eb,Moch:2004pa,Vogt:2004mw,Moch:2004xu,Vermaseren:2005qc,Vogt:2006bt,Moch:2007rq,Moch:2008fj,Davies:2016ruz,Blumlein:2021enk,Blumlein:2022gpp}.\footnote{We
notice that, by convention, achieving N$^3$LO would in principle imply
computing the anomalous dimensions responsible for the energy
evolution of strong coupling and PDFs to the same relative accuracy as
the coefficient functions. We will address this point below.}  Very
recently the code \ttt{yadism}~\cite{Candido:2024rkr}, which computes
DIS structure functions at this order, became available. However,
their implementation was directly compared to \apfel{} and therefore
relies implicitly on the benchmark presented here.

In this paper, we present a comparison of the structure functions
obtained using the two publicly available codes
\apfel{}~\cite{Bertone:2013vaa,Bertone:2017gds} and
\hoppet{}~\cite{Salam:2008qg}.\footnote{Technically speaking, the
  $\mathcal{O}(\as^3)$ structure functions were already available in
  the {\tt struct-func-devel} branch of \hoppet{} as they were used in
  the {\tt proVBFH}
  code~\cite{Cacciari:2015jma,Dreyer:2016oyx,Dreyer:2018qbw,Dreyer:2018rfu}. However,
  while performing this benchmark, some bugs were found in the
  $\mathcal{O}(\as^3)$ neutral-current structure functions, and it is
  therefore fair to say that they have only been publicly available in
  \hoppet{} as of the {\tt v1.3.0} release, which will be made public
  in a forthcoming paper~\cite{hoppetv130}, but is already available
  on the \hoppet{} GitHub repository. As far as \apfel{} is concerned,
  the $\mathcal{O}(\as^3)$ structure functions are available in the
  current {\tt master} branch of the GitHub repository, as well as in
  {\tt v4.8.0}.}
The benchmark that we present here serves two purposes.
First, it validates the correctness of the structure functions as
implemented in the two programs.
This is a highly non-trivial check in that, although the coefficient
functions are identical in the two programs, the underlying
technologies adopted by the two codes are not.
Secondly, the results presented here provide a reference for any
future numerical implementation of the DIS structure functions.
To make the benchmark as resilient towards the future as possible, we
carry out the benchmark using the same PDF initial conditions as those
used in Ref.~\cite{Giele:2002hx}.
This avoids complications caused by numerical artefacts related to
pre-computed interpolation grids such as those released through the
{\tt LHAPDF} interface~\cite{Buckley:2014ana}. Moreover, this
guarantees the independence of the benchmark from the availability of
a specific PDF set.
The benchmark is carried out both on the single structure functions
and on the reduced cross sections, which is what is often measured in
experiments.

The paper is structured as follows.
In Sect.~\ref{sec:dis-sf}, we review the DIS process and define the
structure functions.
The numerical setup of the benchmark is presented in
Sect.~\ref{sec:setup}.
We finally present the benchmark in Sect.~\ref{sec:benchmark} before
concluding in Sect.~\ref{sec:conclusion}. \ref{app:exact-vs-param}
contains some comments on the large-$y$ behaviour of the non-singlet
N$^3$LO coefficient functions, while \ref{app:bench_tables} presents
look-up benchmark tables for all of the structure functions at NLO,
NNLO, and N$^3$LO accuracy for different values of $Q$ and for Bjorken
$x_{\rm B}\in [10^{-5}:0.9]$.

\section{The DIS structure functions}
\label{sec:dis-sf}

Let us begin by recalling the kinematics of the DIS process. This
process is the inclusive scattering of a lepton $\ell$ with momentum
$k_i$ off a proton $p$ with momentum $P$ via the exchange of a virtual
electroweak gauge boson $V$ with momentum $q$ and large (negative)
virtuality $Q^2=-q^2\gg\Lambda_{\rm QCD}^2$, where $\Lambda_{\rm QCD}$
is the typical hadronic scale. Due to the large virtuality of the
vector boson, the proton breaks up leaving in the final state the
scattered lepton $\ell'$ with momentum $k_f=k_i-q$ and a remnant $X$,
with respect to which we are fully inclusive:
\begin{equation}
\ell(k_i)+p(P)\rightarrow V(q) \rightarrow \ell'(k_f)+X\,.
\end{equation}
It is useful to introduce the customary DIS variables $\xdis$
(Bjorken's variable) and $\ydis$ (inelasticity) defined as:
\begin{align}
  \xdis = \frac{Q^2}{2 P \cdot q},
  \qquad  \ydis = \frac{P \cdot q}{P \cdot k_i}=\frac{Q^2}{\xdis s}\,,
  \label{eq:dis-variables}
\end{align}
where $s=(k_i+P)^2$ is the collision center-of-mass energy
squared.\footnote{We assume that incoming lepton and proton are both
  massless, \textit{i.e.} $k_i^2=P^2=0$.} In order to describe the
interaction between the proton and the appropriate electroweak current,
\textit{i.e.} the neutral current (NC) mediated by a $V=\gamma/Z$
boson or the charged current (CC) mediated by a $V=W^\pm$ boson, it is
useful to consider the hadronic tensor $W_{\mu\nu}^V$. The
spin-averaged hadronic tensor defines the structure functions $F_1$,
$F_2$, and $F_3$ through~\cite{ParticleDataGroup:2022pth}:
\begin{align}
  W^V_{\mu\nu} = \left( -g_{\mu\nu} + \frac{q_\mu q_\nu}{q^2}\right)F^V_1(\xdis,Q^2) + \frac{\hat{P}_\mu\hat{P}_\nu}{P\cdot q}F^V_2(\xdis,Q^2) - i\epsilon_{\mu\nu\alpha\beta}\frac{q^\alpha P^\beta}{2P\cdot q}F^V_3(\xdis,Q^2),
\end{align}
where:
\begin{align}
  \hat{P}_\mu = P_\mu - \frac{P\cdot q}{q^2}q_\mu\,.
\end{align}
It is also customary to define the longitudinal structure function as
$F_L^V = F_2^V - 2\xdis F_1^V$.
With the structure functions at hand, we can write the NC cross
section ($\ell^\pm p \to \ell^\pm + X$) as:
\begin{equation}
\frac{\dd\sigma_{\NC}^\pm}{\dd \xdis \dd Q^2} =
\frac{2\pi\alpha^2}{\xdis Q^4}y_+
\left[ F_2^{\gamma/Z} \mp \frac{y_-}{y_+} \xdis F_3^{\gamma/Z}  - \frac{y^2}{y_+} F_L^{\gamma/Z} \right]\,,
\label{eq:NCsigma}
\end{equation}
where $y_\pm=1\pm(1-y)^2$ and $\alpha$ is the fine structure constant.
Similarly, the CC cross section for
($\ell^\pm p \to \nu_\ell (\overline{\nu}_\ell) + X$) reads:
\begin{equation}
\frac{\dd\sigma_{\CC}^\pm}{\dd \xdis \dd Q^2} =
\frac{\pi\alpha^2}{8 \xdis\sin^4\theta_W }\left(\frac{1}{M_W^2 +
    Q^2}\right)^2 \left[y_+ F_2^{W^{\pm}} \mp y_- \xdis F_3^{W^{\pm}} - y^2
  F_L^{W^{\pm}}\right],\,
\label{eq:CCsigma}
\end{equation}
where $M_W$ is the mass of the $W$ boson and $\theta_W$ is the
electroweak mixing angle. Some experiments, such as those at the HERA
collider~\cite{H1:2015ubc}, release the so-called ``reduced'' cross
sections that are related to the standard cross sections as follows:
\begin{equation}
  \label{eq:redxsec}
  \begin{array}{l}
    \displaystyle \sigma_{\rm NC, red}^{\pm} = \left[\frac{2\pi\alpha^2}{\xdis
    Q^4}y_+\right]^{-1}\frac{\dd\sigma_{\NC}^\pm}{\dd \xdis \dd
    Q^2}=F_2^{\gamma/Z} \mp \frac{y_-}{y_+} \xdis F_3^{\gamma/Z}  - \frac{y^2}{y_+} F_L^{\gamma/Z}\,,\\
    \\
       \displaystyle  \sigma_{\rm CC, red}^{\pm} = \left[\frac{\pi\alpha^2}{4\sin^4\theta_W \xdis}\left(\frac{1}{M_W^2 +
    Q^2}\right)^2\right]^{-1}\frac{\dd\sigma_{\CC}^\pm}{\dd \xdis \dd
    Q^2}=\frac{y_+}{2} F_2^{W^{\pm}} \mp \frac{y_-}{2} \xdis F_3^{W^{\pm}} - \frac{y^2}{2}
  F_L^{W^{\pm}}\,.
  \end{array}
\end{equation}

The QCD collinear factorisation theorem allows us to express the
structure functions as convolutions of the PDFs, $f_a$, with the
short-distance coefficient functions, $C_i$:\footnote{Note that,
according to the usual definitions, the overall factor $x_{\rm B}$ on
the r.h.s. of Eq.~(\ref{eq:conv-structf}) only applies to $F_2^V$ and
$F_L^V$, while it is not present for $F_3^V$. This is the meaning of
the parentheses around $x_{\rm B}$ itself and the indices
$i=(2),(L)$.}
\begin{equation}
  \label{eq:conv-structf}
  F_i^V\equiv F_i^V(x_{\rm B}, Q^2) = (x_{\rm B})\sum_{a} \left[C_{i,a}^{V} \otimes f_a
  \right]\left(x_{\rm B},Q^2,\mu_{\rm R}^2,\mu_{\rm F}^2\right)\,,\quad i=(2),(L),3\,,
\end{equation}
where the index $a$ runs over the gluon ($a=g$) and all active quark
flavours and anti-flavours ($a=q,\overline{q}$) at the scale
$Q^2$.\footnote{Specifically, if the mass $m_q$ of the quark flavour
  $q$ is such that $m_q^2<Q^2$, this flavour contributes to the cross
  section, otherwise it does not. In practice, down, up, and strange
  quarks always contribute in that their masses are always far below
  the typical hard scale $Q^2$ where factorisation applies. For this
  reason they are called light quarks. Conversely, charm, bottom, and
  possibly top quarks get activated at the respective mass scales,
  therefore they are referred to as heavy quarks. This is a possible
  implementation of the of the so-called \textit{decoupling
    theorem}~\cite{Appelquist:1974tg} that goes under the name of
  zero-mass variable-flavour-number scheme (ZM-VFNS). According to
  this theorem, for $m_q^2\ll Q^2$ the quark flavour $q$ must drop
  from the calculation. The ZM-VFNS enforces this constraint already
  when $m_q^2<Q^2$, which effectively amounts to neglecting positive
  powers of the ratio $m_q^2/Q^2$.} The Mellin-convolution symbol
$\otimes$ implies one of the following equivalent integrals:
\begin{equation}
  \begin{array}{rcl}
  \displaystyle \left[C_{i,a}^{V} \otimes f_a\right]\left(x_{\rm B},Q^2,\mu_{\rm R}^2,\mu_{\rm F}^2\right) &=&\displaystyle  \int_0^1dy\int_0^1dz\,
    \delta(x_{\rm B}-yz)C_{i,a}^V\left(y,\alpha_s(\mu_{\rm
    R}),\frac{\mu_{\rm R}^2}{Q^2},\frac{\mu_{\rm
                              F}^2}{Q^2}\right)f_a(z,\mu_{\rm F})\\
    \\
    &=&\displaystyle \int_{x_{\rm B}}^1
    \frac{dy}{y}C_{i,a}^V\left(y,\alpha_s(\mu_{\rm
    R}),\frac{\mu_{\rm R}^2}{Q^2},\frac{\mu_{\rm
        F}^2}{Q^2}\right)f_a\left(\frac{x_{\rm B}}{y},\mu_{\rm F}\right) \\
    \\
    &=&\displaystyle \int_{x_{\rm B}}^1
    \frac{dz}{z}C_{i,a}^V\left(\frac{x_{\rm B}}{z},\alpha_s(\mu_{\rm
    R}),\frac{\mu_{\rm R}^2}{Q^2},\frac{\mu_{\rm
                              F}^2}{Q^2}\right)f_a(z,\mu_{\rm F})\,.
    \end{array}
  \label{eq:convolution}
\end{equation}

\subsection{Renormalisation and factorisation scale dependence}
\label{sec:ren-fac-dep}

In this section, we derive the explicit dependence of the coefficient
functions on the \textit{arbitrary} renormalisation and factorisation
scales $\mu_{\rm R}$ and $\mu_{\rm F}$, respectively. The coefficient
function $C_{i,a}^{V}$ is an explicit function of the strong coupling
$\alpha_s(\mu_{\rm R})$ that admits the perturbative expansion:
\begin{equation}
C_{i,a}^V\left(y,\alpha_s(\mu_{\rm R}),\frac{\mu_{\rm
      R}^2}{Q^2},\frac{\mu_{\rm
      F}^2}{Q^2}\right)=\sum_{n=0}^{\infty}\left(\frac{\alpha_s(\mu_{\rm
    R})}{4\pi}\right)^n C_{i,a}^{V,[n]}\left(y,\frac{\mu_{\rm
    R}^2}{Q^2},\frac{\mu_{\rm F}^2}{Q^2}\right)\,.
\label{eq:perturbativeseries}
\end{equation}
Under the assumption $\alpha_s(\mu_{\rm R})\ll 1$, the sum on the
r.h.s. can be truncated to order $k$ to obtain a N$^k$LO approximation
of the structure functions.\footnote{If $i=L$, \textit{i.e.} in the
  case of the longitudinal structure function, the contribution $k=0$
  to the series on the r.h.s. of Eq.~(\ref{eq:perturbativeseries}) is
  identically zero. Therefore, in the case of $F_L$ a strictly N$^k$LO
  requires truncating that series at $n=k+1$. This is a consequence of
  the Callan-Gross relation~\cite{Callan:1969uq},
  $F_2^V = 2 \xdis F_1^V$, valid in the parton model for spin-1/2
  particles, that implies that $F_L^V=0$ at
  $\mathcal{O}(\alpha_s^0)$.} $C_{i,a}^{V}$ also depends on the ratios
$\mu_{\rm R}^2/Q^2$ and $\mu_{\rm F}^2/Q^2$. Despite that the scales
$\mu_{\rm R}$ and $\mu_{\rm F}$ are in principle arbitrary, in a
fixed-order calculation where the series in
Eq.~(\ref{eq:perturbativeseries}) includes a finite number of terms,
the presence of logarithms of these ratios requires these scales to be
of order $Q=\sqrt{Q^2}$. In this way, the ratios $\mu_{\rm R}^2/Q^2$
and $\mu_{\rm F}^2/Q^2$ are both of order one and do not compromise
the convergence of the perturbative series. Variations of
$\mu_{\rm R}$ and $\mu_{\rm F}$ around $Q$ by a modest factor,
typically of two, are often used as a proxy to estimate the possible
impact of unknown higher-order corrections. This is due to the fact
that any variation of $\mu_{\rm R}$ and $\mu_{\rm F}$ is compensated
order by order in $\alpha_s$ by the evolution of strong coupling and
PDFs, that in turn obey their own renormalisation group equations
(RGEs):
\begin{equation}
  \begin{array}{rcl}
\displaystyle \frac{d\ln\alpha_s(\mu_{\rm R})}{d\ln\mu_{\rm R}^2}&=&\displaystyle
                                                     \beta(\alpha_s(\mu_{\rm
                                                     R}^2))=-\sum_{n=0}^{\infty}\left(\frac{\alpha_s(\mu_{\rm
                                                     R})}{4\pi}\right)^{n+1}\beta_n\,,\\
    \\
    \displaystyle \frac{df_a(x,\mu_{\rm F})}{d\ln\mu_{\rm F}^2}&=&\displaystyle
                                                     \sum_{b}[P_{ab}\otimes f_b](x,\mu_{\rm F})=\sum_{b}\sum_{n=0}^{\infty}\left(\frac{\alpha_s(\mu_{\rm
                                                     F})}{4\pi}\right)^{n+1}[P_{ab}^{[n]}\otimes f_b](x,\mu_{\rm F})\,.\\
  \end{array}
  \label{eq:RGEs}
\end{equation}
The RGEs for PDFs are usually referred to as DGLAP
equations~\cite{Lipatov:1974qm,Gribov:1972ri,Altarelli:1977zs,Dokshitzer:1977sg}
and the kernels $P_{ab}$ are called splitting functions. Given the
appropriate set of boundary conditions, the solutions of these RGEs
determine the evolution (or running) of both $\alpha_s$ and $f_a$ to
any scale. Similarly to the DIS coefficient functions, the anomalous
dimensions $\beta$ and $P_{ab}$ are expandable in powers of
$\alpha_s$.

When computing a DIS structure function at N$^k$LO accuracy, it is
customary to truncate also the perturbative expansions in
Eq.~(\ref{eq:RGEs}) at the same relative order $k$. However, this is
mostly a conventional procedure that is not strictly mandatory for a
correct counting of the perturbative accuracy.  Strictly speaking, the
truncation of the expansion in Eq.~(\ref{eq:perturbativeseries}) is
responsible for the \textit{fixed-order} accuracy, while the
truncation of the expansions Eq.~(\ref{eq:RGEs}) defines the
\textit{logarithmic} accuracy. The fixed-order accuracy counts how
many corrections proportional to an integer non-negative power of
$\alpha_s$ are included \textit{exactly} in the coefficient
functions. In the DIS case, the $\mathcal{O}(\alpha_s^0)$ contribution
gives leading-order (LO) accuracy, the inclusion of the
$\mathcal{O}(\alpha_s)$ corrections gives next-to-leading-order (NLO)
accuracy, and so on. The logarithmic accuracy instead counts the
number of all-order towers of logarithms of the kind $\alpha_s^mL^n$,
with $L=\ln\left(\mu_{\rm R,F}/Q_0\right)$ and $Q_0$ the
boundary-condition scale, that are being resummed by means of the
evolution of $\alpha_s$ and PDFs.\footnote{Note that the boundary
  scale $Q_0$ can, and often is, different for $\alpha_s$ and PDFs.}
Leading-logarithmic (LL) accuracy is achieved summing all
$\alpha_s^nL^n$ terms, next-to-leading-logarithmic (NLL) accuracy
requires summing all $\alpha_s^{n+1}L^n$ terms, and so on. It is worth
noting that the distinction between fixed-order and logarithmic
counting is effective only when $\alpha_sL \sim 1$, \textit{i.e.} when
$L$ is large enough to compensate for the assumed smallness of
$\alpha_s$. This holds when $\mu_{\rm R,F}\gg Q_0$ (or
$\mu_{\rm R,F}\ll Q_0$), which is often the case in phenomenological
applications.

Although fixed-order and logarithmic accuracies have two different
origins, they are not entirely unrelated. Indeed, loosely speaking,
the summation of logarithms also contributes to the fixed-order
counting; for instance the term $\alpha_sL$, which belongs to the LL
tower, can also be regarded as a NLO contribution to the DIS structure
functions. Therefore, NLO accuracy must come with \textit{at least} LL
resummation. In general, N$^k$LO accuracy for the DIS coefficient
functions requires a minimal resummation accuracy of
N$^{k-1}$LL. However, it is not incorrect to match N$^k$LO coefficient
functions to N$^{k}$LL $\alpha_s$ and PDF resummation: this is what
conventional N$^k$LO accuracy for structure functions prescribes.

In the following, we will adopt the ``conventional'' counting for the
computation of the DIS structure functions up to NNLO, \textit{i.e.}
we will match N$^k$LO coefficient functions to N$^{k}$LL resummation,
with $k=0,1,2$. At N$^3$LO, we will instead use the ``minimal''
prescription and match N$^3$LO coefficient functions to NNLL
resummation. The reason for this choice is that, as of today,
splitting functions are fully known only up to
$\mathcal{O}(\as^3)$~\cite{Gross:1973ju,Georgi:1974wnj,Floratos:1977au,Floratos:1978ny,Gonzalez-Arroyo:1979guc,Gonzalez-Arroyo:1979qht,Curci:1980uw,Furmanski:1980cm,Floratos:1981hs,Hamberg:1991qt,Moch:2004pa,Vogt:2004mw,Blumlein:2021enk}.
Accompanied by the $\mathcal{O}(\as^3)$ corrections to the
$\beta$-function in
Eq.~(\ref{eq:RGEs})~\cite{Politzer:1973fx,Gross:1973id,Caswell:1974gg,Tarasov:1980au,Larin:1993tp}
and the mass threshold corrections to both the running coupling and
the parton
distributions~\cite{Chetyrkin:1997sg,Buza:1995ie,Bierenbaum:2007qe},
this allows us to achieve plain NNLL resummation. While the
$\mathcal{O}(\as^4)$ corrections to the $\beta$-function are
known~\cite{vanRitbergen:1997va,Czakon:2004bu}, this is not the case
for the splitting functions, which prevents attaining exact N$^3$LL
resummation, in spite of the recent significant progress made in
determining them at this
order~\cite{Moch:2021qrk,Falcioni:2023luc,Falcioni:2023vqq,Gehrmann:2023cqm,Falcioni:2023tzp,Moch:2023tdj,Gehrmann:2023iah,Falcioni:2024xyt}. In
contrast, all mass threshold corrections to the running
coupling~\cite{Chetyrkin:1997sg} and the parton distributions are known
at this
order~\cite{Ablinger:2010ty,Blumlein:2012vq,ABLINGER2014263,Ablinger:2014nga,Ablinger:2014vwa,Behring:2014eya,Ablinger:2019etw,Behring:2021asx,Ablinger:2023ahe,Ablinger:2024xtt}.

It is also worth mentioning that, in order to compute the single
ingredients of the factorisation formula for the structure functions
on the r.h.s. of Eq.~(\ref{eq:conv-structf}), it is necessary to
specify a renormalisation/factorisation scheme. While the dependence
on the scheme cancels out order by order in $\alpha_s$, it determines
the specific form of the coefficient functions and of the anomalous
dimensions ($\beta$-function and splitting functions). Throughout this
work, we will use the modified minimal-subtraction
($\overline{\mbox{MS}}$) scheme.

The evolution equations in Eq.~(\ref{eq:RGEs}) allow us to determine
the dependence on $\mu_{\rm R}$ and $\mu_{\rm F}$ of the perturbative
coefficients $C_{i,a}^{V,[n]}$ in
Eq.~(\ref{eq:perturbativeseries}). Indeed, provided that
$\mu_{\rm R,F}\simeq Q$, one can perturbatively solve the RGEs in
Eq.~(\ref{eq:RGEs}) to evolve $\alpha_s$ and PDFs from the scales
$\mu_{\rm R}$ and $\mu_{\rm F}$, respectively, to the scale
$Q$~\cite{Furmanski:1981cw,vanNeerven:2000uj,Buehler:2013fha},
obtaining:
\begin{equation}
  \label{eq:as-running}
  \frac{\as(Q)}{4\pi} = \frac{\as(\muR)}{4\pi} + \left(\frac{\as(\muR)}{4\pi}\right)^2 L_{R}\beta_0 
  + \left(\frac{\as(\muR)}{4\pi}\right)^3 (L_{R}^2\beta_0^2 + L_{R}\beta_1) + \mathcal{O}(\as^4)\,,
\end{equation}
and:
\begin{equation}
  \label{eq:pdf-expansion}
  \begin{array}{rcl}
  f_a(x,Q) & =&\displaystyle f_a(x, \muF) - L_{F} \Bigg\{\left(\frac{\as(\muR)}{4\pi}\right) P_{ab}^{[0]}
        \\
    \\
  &  +&\displaystyle \left(\frac{\as(\muR)}{4\pi}\right)^2  \Bigg[
  P_{ab}^{[1]} - \frac12 L_{F} P_{ac}^{[0]}\otimes P_{cb}^{[0]}
  - \left(\frac{L_{F}}{2} - L_{R}\right) \beta_0 P_{ab}^{[0]} \Bigg]
    \\
    \\
  & +&\displaystyle \left(\frac{\as(\muR)}{4\pi}\right)^3  \Bigg[
  P_{ab}^{[2]} - \frac12 L_{F} \left(P_{ac}^{[0]} \otimes P_{cb}^{[1]} + P_{ac}^{[1]}\otimes
       P_{cb}^{[0]}\right) + \frac16 L_{F}^2 P_{ac}^{[0]}\otimes P_{cd}^{[0]}\otimes P_{db}^{[0]}  \\
    \\
  & +&\displaystyle  \left(\frac{L_{F}}{2} - L_{R}\right) \beta_0
       \left(L_{F} P_{ac}^{[0]}\otimes P_{cb}^{[0]}- 2 P_{ab}^{[1]}\right)   + \left(L_{R}^2 - L_{F} L_{R} +
      \frac13  L_{F}^2\right) \beta_0^2  P_{ab}^{[0]}  \\
    \\
  &-&\displaystyle \left(\frac{L_{F}}{2} - L_{R}\right) \beta_1 P_{ab}^{[0]}
    \Bigg] \Bigg\}\otimes f_b(x, \muF) +\mathcal{O}(\alpha_s^4)\,,
    \end{array}
\end{equation}
where a summation over repeated indices is understood and we
introduced the shorthand notation:
\begin{equation}
  \label{eq:LRQ-notation}
  L_{R} = \ln\left(\frac{\mu_{\rm R}^2}{Q^2}\right)\,,\quad
  L_{F} = \ln\left(\frac{\mu_{\rm F}^2}{Q^2}\right)\,.
\end{equation}
With these equalities at hand, one can solve iteratively order by
order in $\alpha_s$ the following equality:
\begin{equation}
  \sum_{a} \left[C_{i,a}^{V} \otimes f_a
\right]\left(x_{\rm B},Q^2,\mu_{\rm R}^2,\mu_{\rm F}^2\right)=\sum_{a} \left[C_{i,a}^{V} \otimes f_a
\right]\left(x_{\rm B},Q^2,Q^2,Q^2\right)\,,
\end{equation}
which immediately implies:
\begin{equation}
  \small
  \begin{array}{rcl}
    \displaystyle C_{i,a}^{V,[0]}\left(y,\frac{\mu_{\rm
    R}^2}{Q^2},\frac{\mu_{\rm F}^2}{Q^2}\right) &=&\displaystyle
                                                    c_{i,a}^{V,[0]}(y)\,,\\
    \\
    \displaystyle C_{i,a}^{V,[1]}\left(y,\frac{\mu_{\rm
    R}^2}{Q^2},\frac{\mu_{\rm F}^2}{Q^2}\right) &=&\displaystyle
                                                    c_{i,a}^{V,[1]}(y)-L_F
                                                    \left[c_{i,b}^{V,[0]}\otimes
                                                    P_{ba}^{[0]}\right](y)\,,\\
    \\
        \displaystyle C_{i,a}^{V,[2]}\left(y,\frac{\mu_{\rm
    R}^2}{Q^2},\frac{\mu_{\rm F}^2}{Q^2}\right) &=&\displaystyle
                                                    c_{i,a}^{V,[2]}(y)+L_R\beta_0 c_{i,a}^{V,[1]}(y) -L_F\left[c_{i,b}^{V,[1]}
                                                    \otimes P_{ba}^{[0]}\right](y)\\
    \\
    &+&\displaystyle L_F c_{i,b}^{V,[0]}\otimes
                                                    \left[\left(\frac{L_F}{2}-
                                                    L_R\right)\beta_0P_{ba}^{[0]}+\frac12L_FP_{bc}^{[0]}\otimes
                                                    P_{ca}^{[0]}-
                                                    P_{ba}^{[1]}\right](y) \,,\\
    \\
        \displaystyle C_{i,a}^{V,[3]}\left(y,\frac{\mu_{\rm
    R}^2}{Q^2},\frac{\mu_{\rm F}^2}{Q^2}\right) &=&\displaystyle
                                                    c_{i,a}^{V,[3]}(y)
                                                    +2 L_R \beta_0
                                                    c_{i,a}^{V,[2]}(y)
                                                    -L_F
                                                    \left[c_{i,b}^{V,[2]}\otimes
                                                    P_{ba}^{[0]}\right](y)+\left(L_R^2\beta_0^2+ L_R\beta_1\right) c_{i,a}^{V,[1]}(y) \\
    \\
    &+&\displaystyle L_F c_{i,b}^{V,[1]}\otimes
                                                    \left[\left(\frac{L_F}{2}-2 L_R\right)
                                                    \beta_0 
                                                    P_{ba}^{[0]}+\frac{L_F}{2}
                                                    P_{bc}^{[0]}\otimes
                                                    P_{ca}^{[0]}-P_{ba}^{[1]}\right](y)\\
    \\
    &+&\displaystyle L_F c_{i,b}^{V,[0]}\otimes
                                                    
                                                    \bigg[-\beta_0^2
                                                    P_{ba}^{[0]}
                                                    \left(\frac{L_F^2}{3}-L_FL_R+L_R^2\right)\\
    \\
    &-&\displaystyle \left(\frac{L_F}{2}-L_R\right)
                                                    \beta_0 (L_F
                                                    P_{bc}^{[0]}\otimes
                                                    P_{ca}^{[0]}-2
                                                    P_{ba}^{[1]})-\frac{L_F^2}{6}P_{bc}^{[0]}\otimes
                                                    P_{cd}^{[0]}\otimes
        P_{da}^{[0]}\\
    \\
    &+&\displaystyle \frac{L_F}{2}(P_{bc}^{[0]}\otimes
                                                    P_{ca}^{[1]}+P_{bc}^{[1]}\otimes
        P_{ca}^{[0]})-P_{ba}^{[2]}\bigg](y)\,,
  \end{array}
  \label{eq:CFexppansion}
\end{equation}
where we have defined:
\begin{equation}
c_{i,a}^{V,[n]}(y) = C_{i,a}^{V,[n]}\left(y,1,1\right)\,.
\end{equation}
Eq.~(\ref{eq:CFexppansion}) allows us to compute the structure
functions up to N$^3$LO accuracy for any choice of renormalisation and
factorisation scales.

\subsection{Characteristics of the structure functions}
\label{sec:strct-fct-char}

We now move to characterising the structure functions. In the NC case,
structure functions have the following structure:
\begin{equation}
  \begin{array}{l}
    \displaystyle F_i^{\gamma/Z} = x_{\rm
    B}\sum_{a}B_a\left[C_{i,\rm NS}^+\otimes f_{a}^{+}+C_{i,\rm
    PS}\otimes f_{\rm PS}+C_{i,g}\otimes f_{g}\right]\,,\quad i=2,L\,,\\
    \\
    \displaystyle F_3^{\gamma/Z}= \sum_{a}D_a\left[C_{3,\rm NS}^-\otimes f_{a}^{-}+C_{3,\rm
    PV}\otimes f_{\rm PV}\right]\,,
  \end{array}
  \label{eq:NCSTDeco}
\end{equation}
where we have defined the following combinations of quark PDFs:
\begin{equation}
  f_{a}^{\pm} = f_a \pm f_{\bar{a}}\,,\qquad f_{\text{PS}} = \sum_{a}f_{a}^{+}\,, \qquad f_{\text{PV}} = \sum_{a}f_{a}^{-}\,.
\end{equation}
Importantly, in the decomposition in Eq.~(\ref{eq:NCSTDeco}) the
coefficient functions are independent of the flavour index
$a$. Conversely, the electroweak charges $B_a$ and $D_a$ do depend on
the flavour index as follows:
\begin{equation}
  \begin{array}{l}
    \displaystyle B_a=e_a^2-2e_aV_\ell V_aP_Z+(V_\ell^2+A_\ell^2)
    (V_a^2+A_a^2)P_Z^2\,,\\
    \\
    \displaystyle D_a=-2e_aA_\ell A_aP_Z+4V_\ell A_\ell V_aA_aP_Z^2\,,
  \end{array}
\end{equation}
where:
\begin{equation}
P_Z=\frac{1}{4\sin^2\theta_{W}\cos^2\theta_{W}}\left(\frac{Q^2}{Q^2+M_Z^2}\right)\,,
\end{equation}
and $\ell$ is the lepton off which the proton scatters. The electric,
vector, and axial charges for quarks and leptons are given in the
Tab.~\ref{tab:couplings}.
\begin{table}[ht]
  \begin{center}
    \begin{tabular}{c|ccc}
      \hline
      $f$ & $e_f$ & $V_f$ & $A_f$  \\
      \hline
      $\vphantom{\displaystyle\frac11}$ $d$, $s$, $b$
          & $-\frac{1}{3}$ & $-\frac{1}{2}+\frac{2}{3}\sin^2\theta_W$
                          & $-\frac{1}{2}$  \\
      $\vphantom{\displaystyle\frac11}$ $u$, $c$, $t$                    & $+\frac{2}{3}$ & $+\frac{1}{2}-\frac{4}{3}\sin^2\theta_W$ & $+\frac{1}{2}$  \\
      \hline
      $\vphantom{\displaystyle\frac11}$ $e$, $\mu$, $\tau$               & $-1$           & $-\frac{1}{2}+2\sin^2\theta_W$           & $-\frac{1}{2}$  \\
      $\vphantom{\displaystyle\frac11}$ $\nu_e$, $\nu_{\mu}$, $\nu_{\tau}$  & $0$            & $+\frac{1}{2}$                           & $+\frac{1}{2}$  \\
      \hline
    \end{tabular}
    \caption{Electric, vector, and axial couplings for up-type,
      down-type, charged leptons, and neutrinos.}\label{tab:couplings}
  \end{center}
\end{table}

We now move to the CC structure functions whose factorised expression
reads:
\begin{equation}
  \begin{array}{l}
    \displaystyle F_i^{W^\pm} = \frac12x_{\rm
  B}\left[\mp C_{i,\rm NS}^-\otimes \delta f_{\rm PV}+(C_{i,\rm NS} + C_{i,\rm PS})\otimes f_{\rm PS}+C_{i,g}\otimes f_{g}\right]\,,\quad i=2,L\,,\\
    \\
    \displaystyle F_3^{W^\pm}= \frac12\left[\pm C_{3,\rm NS}^+\otimes
    \delta f_{\rm PS}+C_{3,\rm
    PV}\otimes f_{\rm PV}\right]\,,
  \end{array}
  \label{eq:CCSTDeco}
\end{equation}
where we have defined the additional quark-PDF combinations:
\begin{equation}
  \delta f_{\rm PS} = \sum_{a\in u\text{-type}} f_a^+ -\sum_{a\in d-\text{type}}  f_a^+\,,\qquad  \delta f_{\rm PV} = \sum_{a\in u\text{-type}} f_a^- -\sum_{a\in d-\text{type}}  f_a^-\,.
\end{equation}
We notice that the expressions in Eq.~(\ref{eq:CCSTDeco}) have been
obtained under the assumption of a CKM matrix equal to the $3\times3$
unity,\footnote{In fact, relying on unitarity, Eq.~(\ref{eq:CCSTDeco})
  is exactly true for any CKM matrix if $n_f=6$.}  which we will also
use in the numerical results presented below. The corresponding
expressions for a generic CKM matrix are considerably more complicated
and we do not present them here.

\section{Numerical setup}
\label{sec:setup}

For our benchmark, rather than relying on a set of tabulated PDFs as
for example delivered by {\tt LHAPDF}~\cite{Buckley:2014ana}, we
decided to use a set of realistic initial-scale conditions having a
simple analytic form and to carry out the evolution ourselves. Besides
the obvious advantage of having full numerical control on our results,
we believe that this choice will allow for easier comparison to our
benchmark results. To this purpose, we selected as initial conditions
for the evolution the parameterisation of Sect.~1.3 of
Ref.~\cite{Giele:2002hx}. Specifically, we chose $Q_0 = \sqrt{2} \GeV$
as an initial scale with $\as(Q_0) = 0.35$. At the initial scale
$Q_0$, only gluon and up, down, and strange quark PDFs are present
while charm, bottom, and top quark PDFs are assumed to be identically
zero and have their production thresholds at
$m_c=(\sqrt{2}+\epsilon)$~GeV,\footnote{The presence of the
  infinitesimal parameter $\epsilon$ in $m_c$ is meant to ensure that
  $Q_0<m_c$ such that the initial conditions for both PDFs and
  $\alpha_s$ are given with $n_f=3$ active flavours. In practice, we
  take $\epsilon = 10^{-9}.$} $m_b= 4.5$~GeV, and $m_t= 175$~GeV,
respectively. At the initial scale $Q_0$, the PDFs are given by:
\begin{subequations}
  \label{eq:init}
  \begin{align}
    x u_v(x, Q_0)   &= 5.107200\,x^{0.8} (1-x)^3\,,\\
    x d_v(x, Q_0)   &= 3.064320\,x^{0.8} (1-x)^4\,,\\
    x\bar d(x, Q_0) &= 0.1939875\,x^{-0.1} (1-x)^6\,,\\
    x\bar u(x, Q_0) &= x\bar d(x) (1-x)\,,\\
    x     s(x, Q_0) &= x\bar s(x, Q_0) = 0.2\,(x\bar d(x, Q_0) + x\bar u(x, Q_0))\,,\\
    x g(x, Q_0) &= 1.7\,x^{-0.1} (1-x)^5\,,
  \end{align}
\end{subequations}
where the valence distributions are defined as $u_v \equiv u - \bar u$
and $d_v \equiv d - \bar d$. We carry out the evolution in the
variable-flavour-number scheme, that is by including quark-mass
thresholds in both coupling and the PDF evolutions. The resulting set
of evolved PDFs both from \apfel{} and \hoppet{} are in perfect
agreement with the tables in Ref.~\cite{Giele:2002hx} at all
perturbative orders. 

Starting from NNLO accuracy, both splitting functions and coefficient
functions become analytically very convoluted and it is customary to
resort to the parameterisations provided by Moch, Vermaseren and
Vogt~\cite{vanNeerven:1999ca,vanNeerven:2000uj,Moch:1999eb,Moch:2004pa,Vogt:2004mw,Moch:2004xu,Vermaseren:2005qc,Vogt:2006bt,Moch:2007rq,Moch:2008fj,Davies:2016ruz}. These
parameterisations are expected to agree with their exact counterparts
at the level of $10^{-4}$ relative accuracy. In this benchmark, we
thus employ exact splitting and coefficient functions up to NLO, while
we use the parameterisations beyond.\footnote{We notice that the exact
expressions are available at all orders in \hoppet{}, as discussed
in~\ref{app:exact-vs-param}.} For the PDF mass threshold corrections
we use the exact expressions at all
orders. In~\ref{app:exact-vs-param}, we comment on the differences
between exact and parametrised coefficient functions.

Finally, we point out that the NC structure functions also depend on
the weak mixing angle (see Sect.~\ref{sec:strct-fct-char}). In this
benchmark, we employ the leading-order relation
$\sin^2\theta_W = 1 - \frac{M_W^2}{M_Z^2}$ to compute it, using
$M_W=80.377$~GeV and $M_Z=91.1876$~GeV for the mass values of $W$ and
$Z$ bosons, respectively.

\section{Benchmark results}
\label{sec:benchmark}

In this section, we present the results of the benchmark between
\apfel{} and \hoppet{}. Here, we will assess the level of agreement
between the two codes at N$^3$LO accuracy by means of a set of
plots. The excellent agreement found at this order (see below)
immediately implies that the agreement at lower orders is at least as
good. In this section, we also take the chance to discuss the impact
of scale variations on DIS structure functions at the available
perturbative orders, as well as the degree of perturbative convergence
moving from LO to N$^3$LO. In~\ref{app:bench_tables}, instead, we
provide look-up tables with predictions at all available perturbative
orders over a broad kinematic range and for all of the DIS structure
functions. We point out that \apfel{} and \hoppet{} are in exact
agreement within the digits shown in those tables.  Therefore, they
can be used as a reference for future numerical implementations of the
DIS structure functions. We also release the code used to produce them
(see~\ref{app:bench_tables} for details).

\begin{figure}[tb!]
  \centering\includegraphics[width=0.32\textwidth]{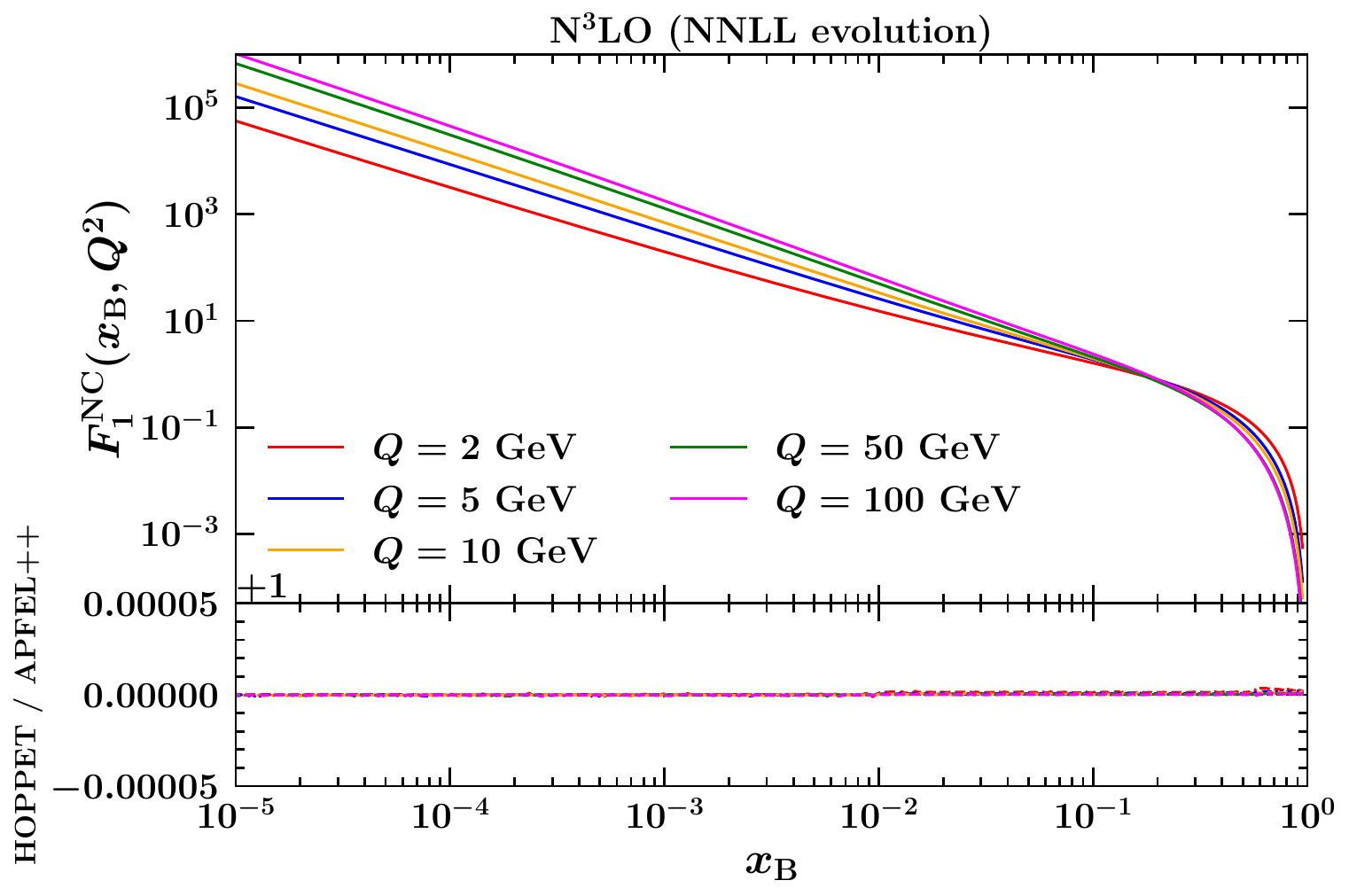}
  \centering\includegraphics[width=0.32\textwidth]{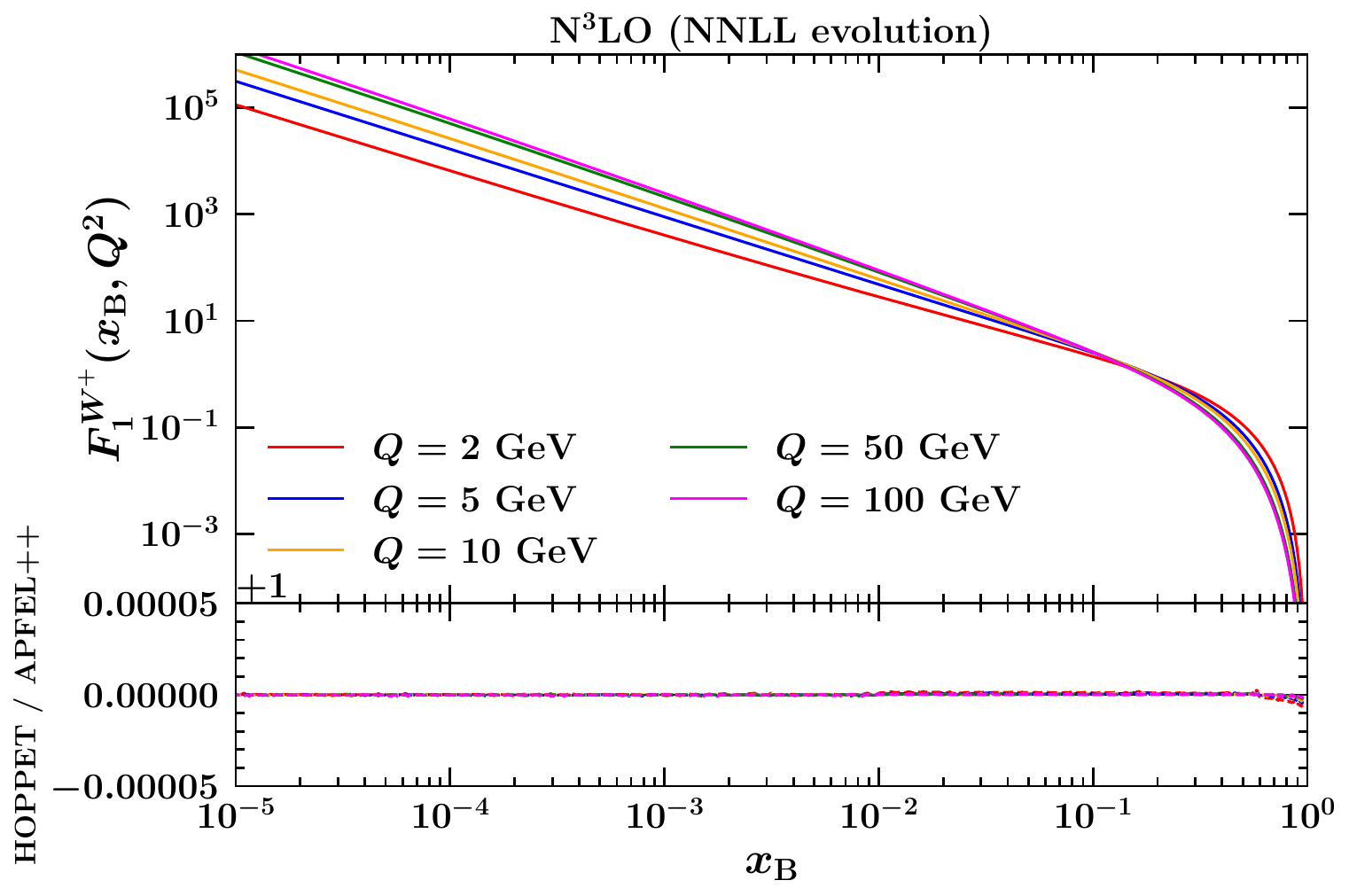}
  \centering\includegraphics[width=0.32\textwidth]{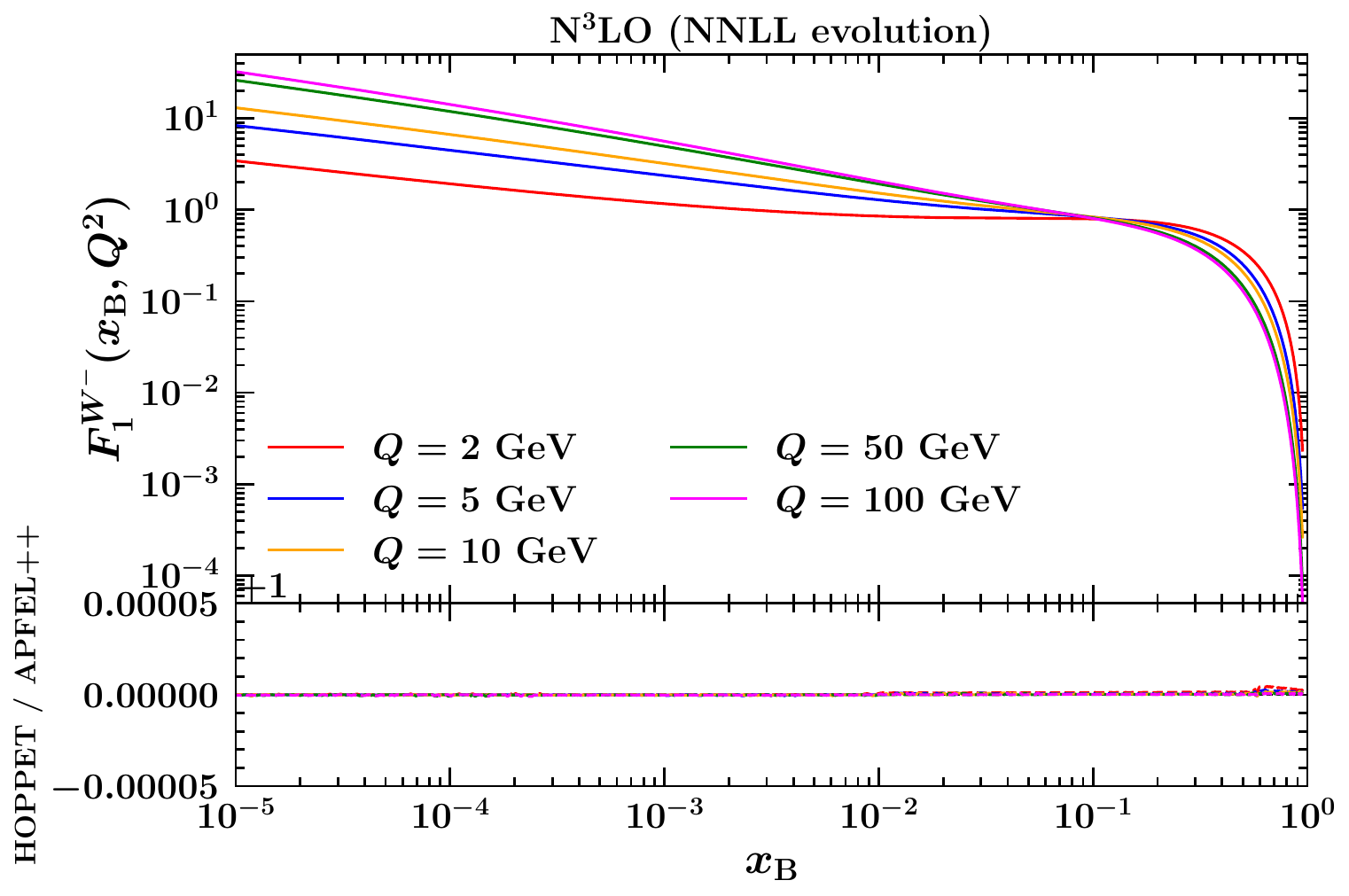}\\
  \centering\includegraphics[width=0.32\textwidth]{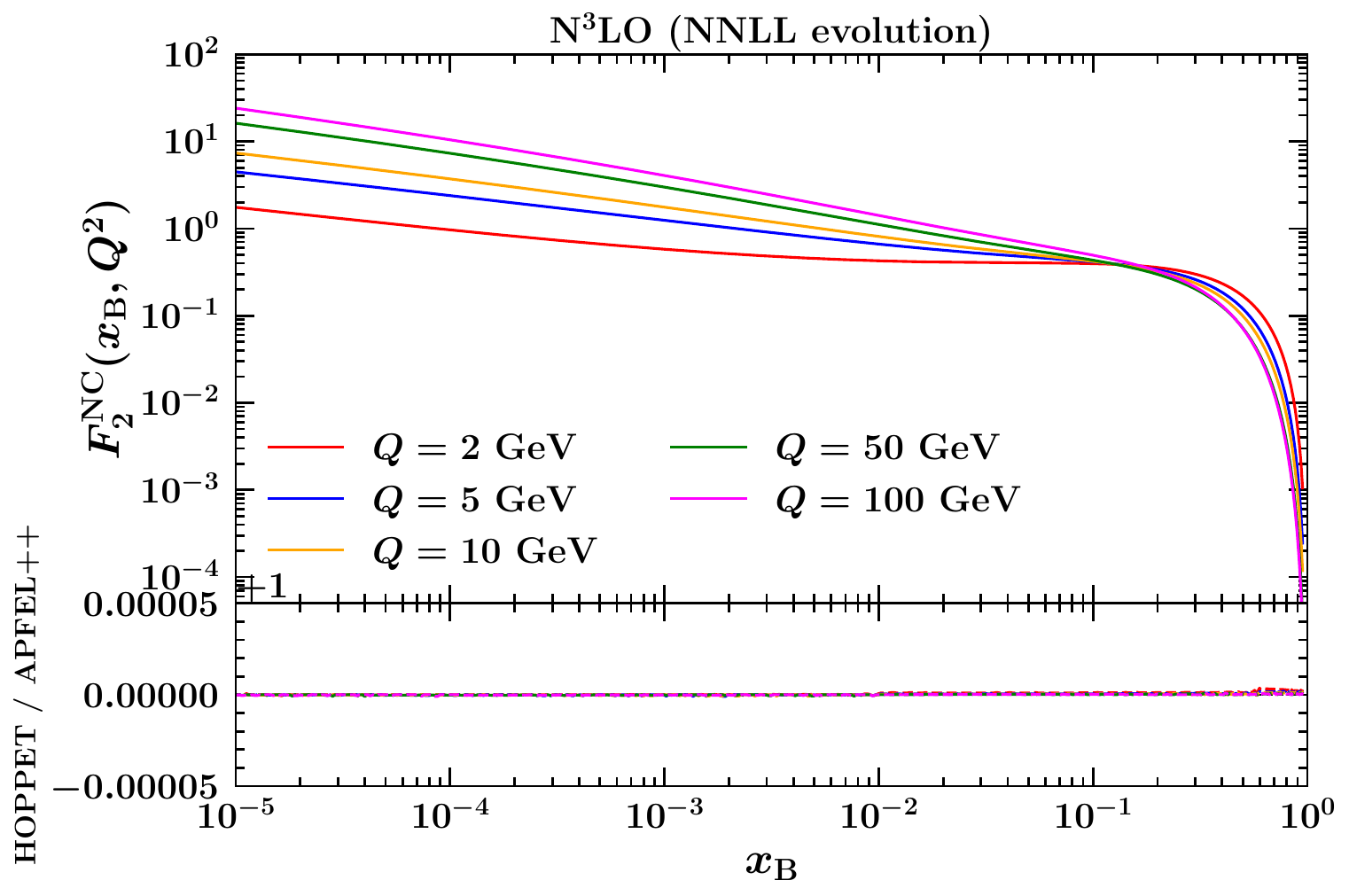}
  \centering\includegraphics[width=0.32\textwidth]{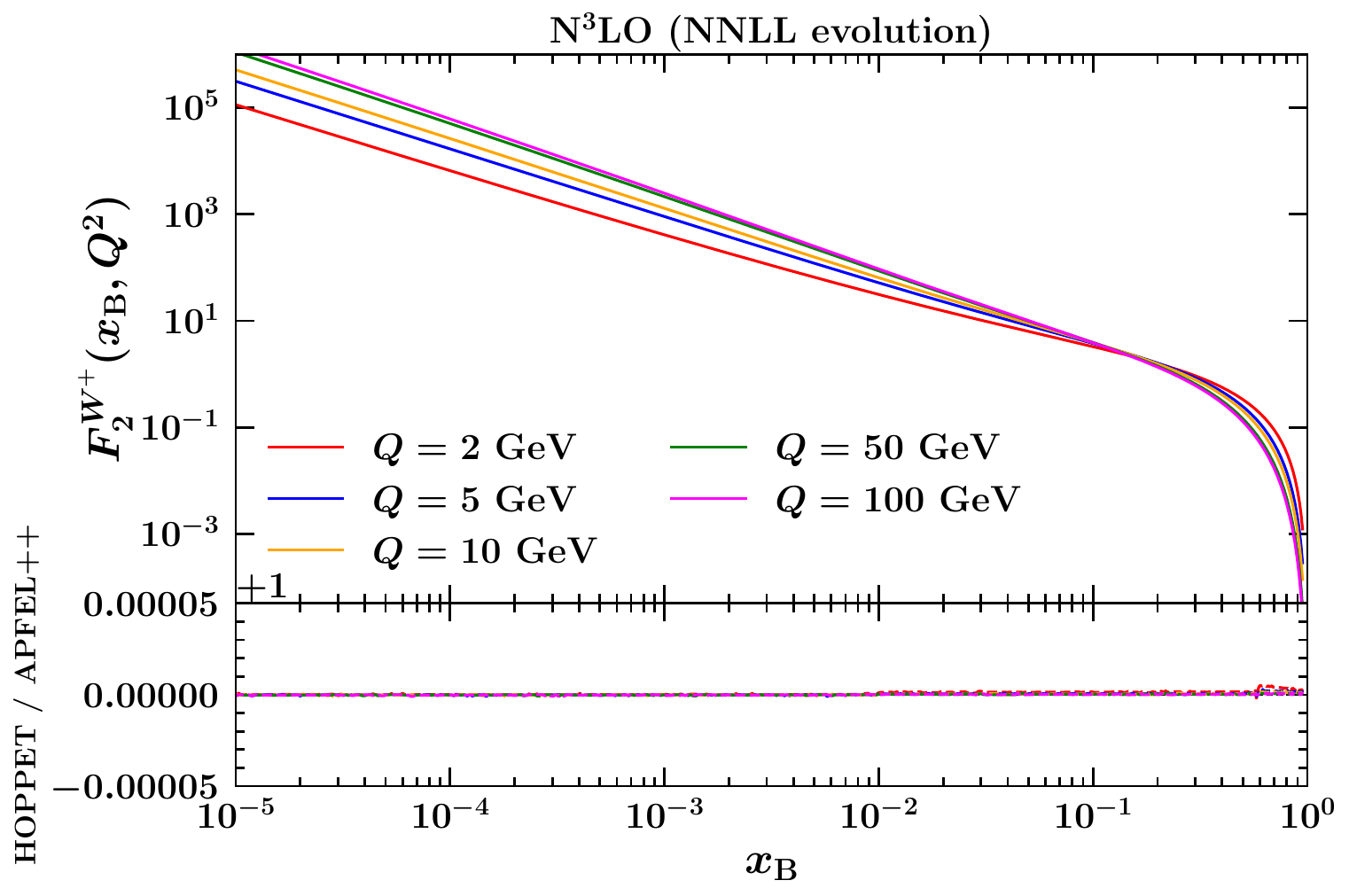}
  \centering\includegraphics[width=0.32\textwidth]{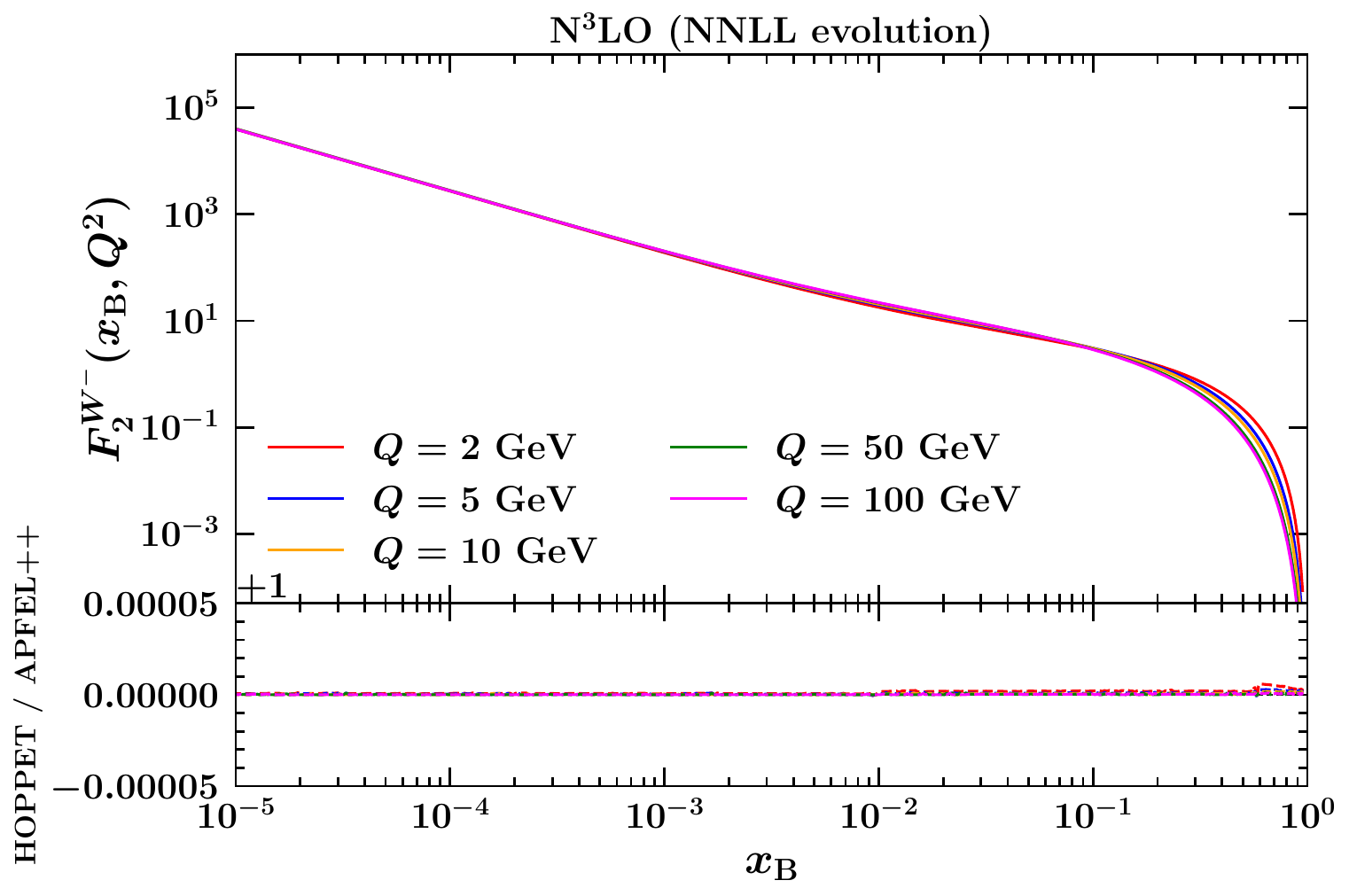}\\
  \centering\includegraphics[width=0.32\textwidth]{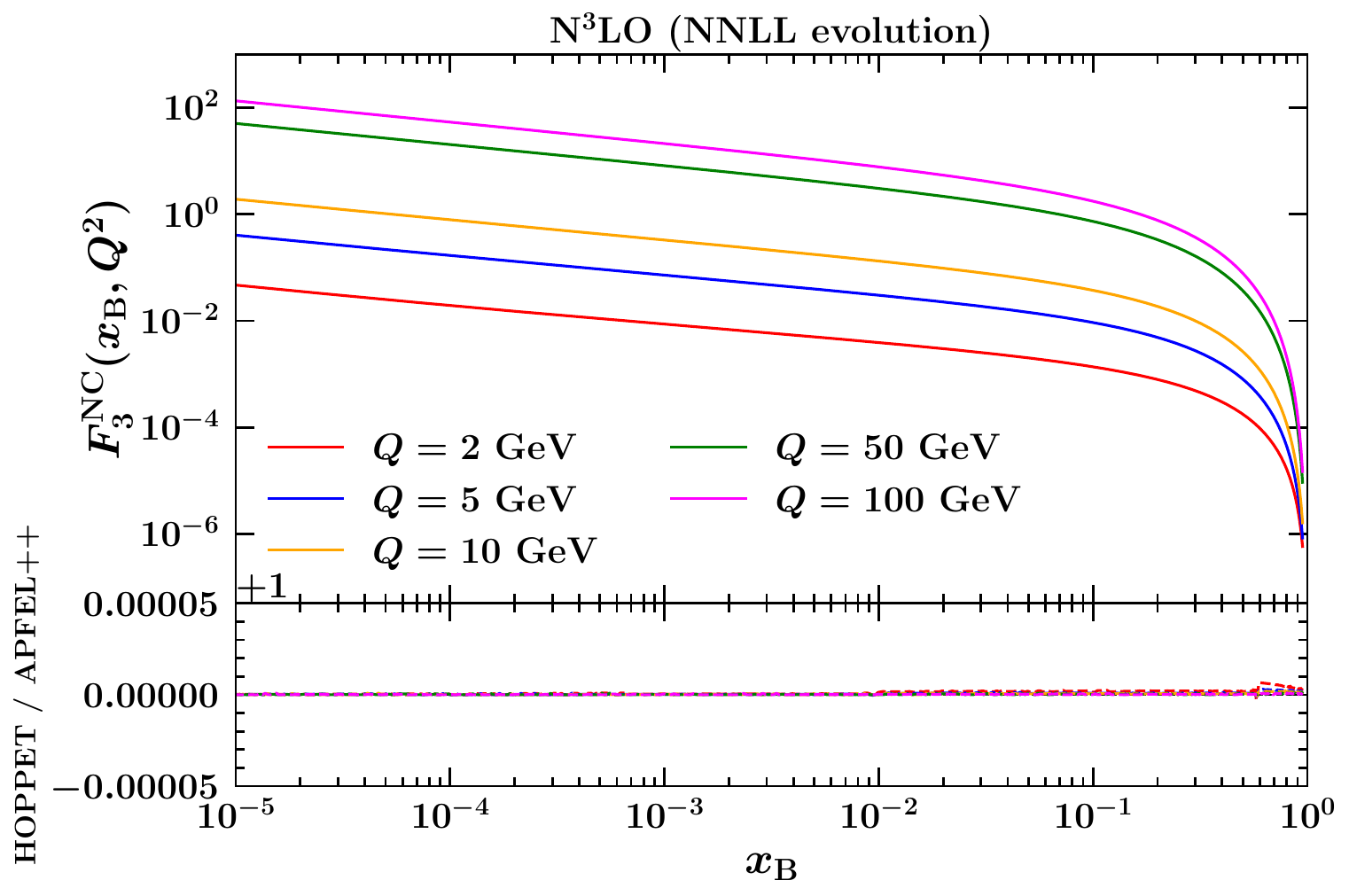}
  \centering\includegraphics[width=0.32\textwidth]{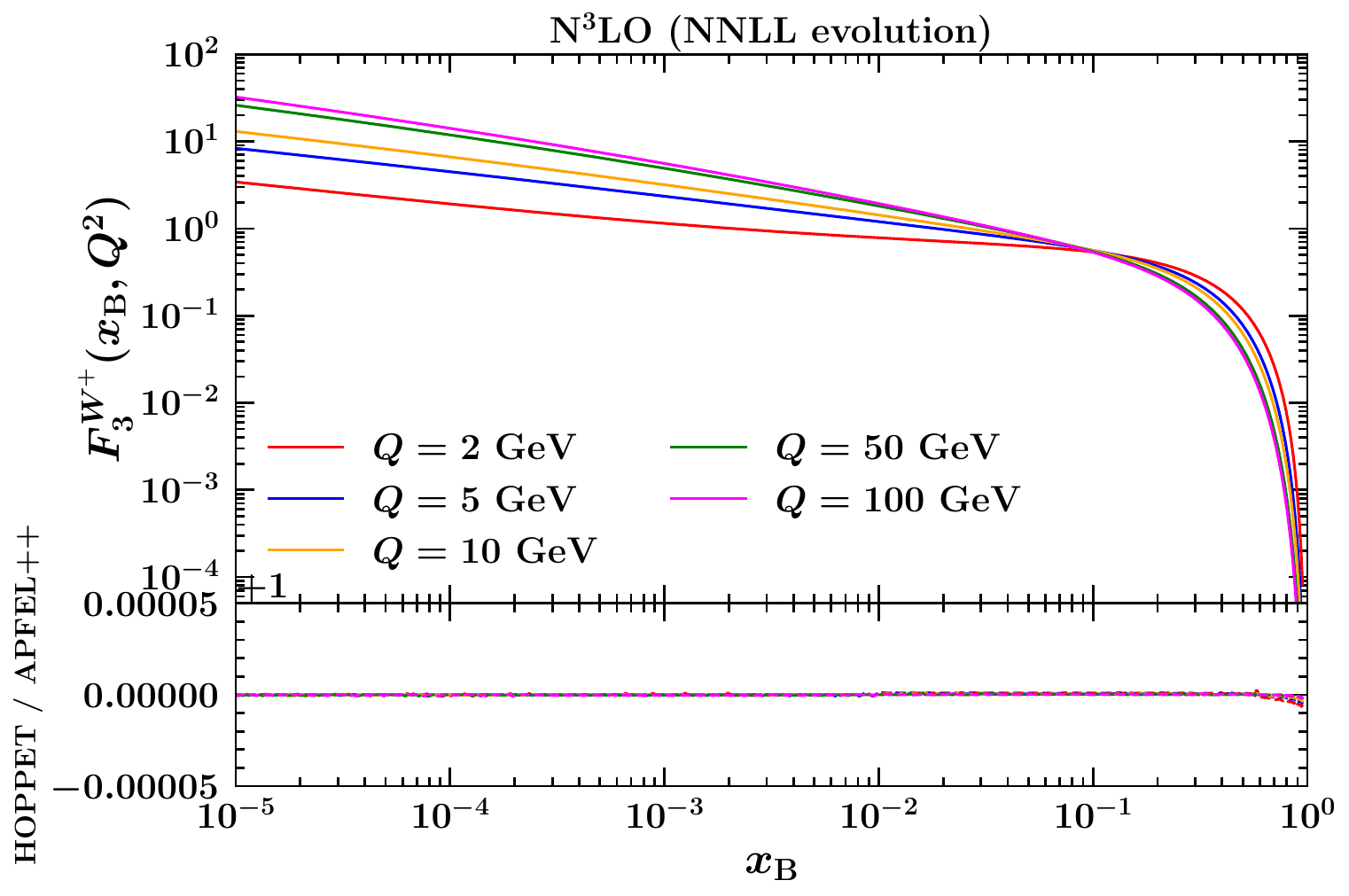}
  \centering\includegraphics[width=0.32\textwidth]{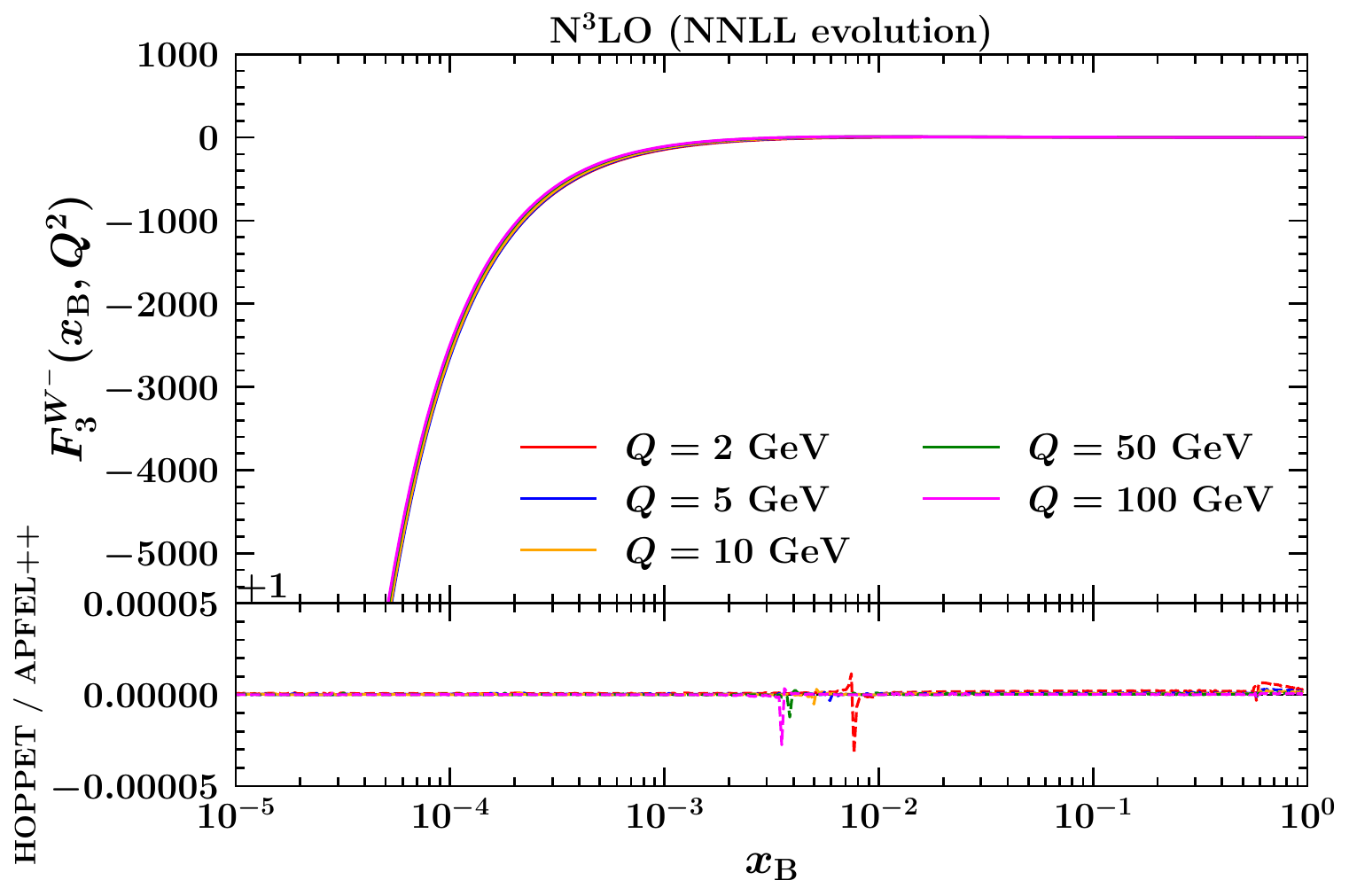}
  \caption{The structure functions $F_1$ (top row), $F_2$ (middle
    row), and $F_3$ (bottom row) for NC (left column), positive CC
    (middle column), and negative CC (right column) at N$^3$LO as
    functions of $x_{\rm B}$ spanning between $10^{-5}$ and $0.9$ and
    for various values of the energy $Q$. The lower panels show the
    ratio between \apfel{} and \hoppet{}.}
  \label{fig:AllStructFuns}
\end{figure}

In Fig.~\ref{fig:AllStructFuns}, we show all structure functions both
in the NC and in the CC channels at N$^3$LO over a wide kinematic
range in $x_{\rm B}$ and $Q$. Here, we set
$\mu_{\rm R}=\mu_{\rm F}=Q$. While the upper panel of each plot
displays the absolute values of the structure functions, the lower
panel shows the ratio between \apfel{} and \hoppet{}. It is evident
that the agreement between the two codes is excellent all across the
board. Specifically, we observe that the relative accuracy is well
below $10^{-5}$ everywhere, except for $F_3^{W^-}$ at around
$x_{\rm B}\sim 3\cdot 10^{-3}-10^{-2}$. However, this slight
degradation in relative accuracy is due to the fact that $F_3^{W^-}$
changes sign in that region.

\begin{figure}[tb!]
  \centering\includegraphics[width=0.49\textwidth]{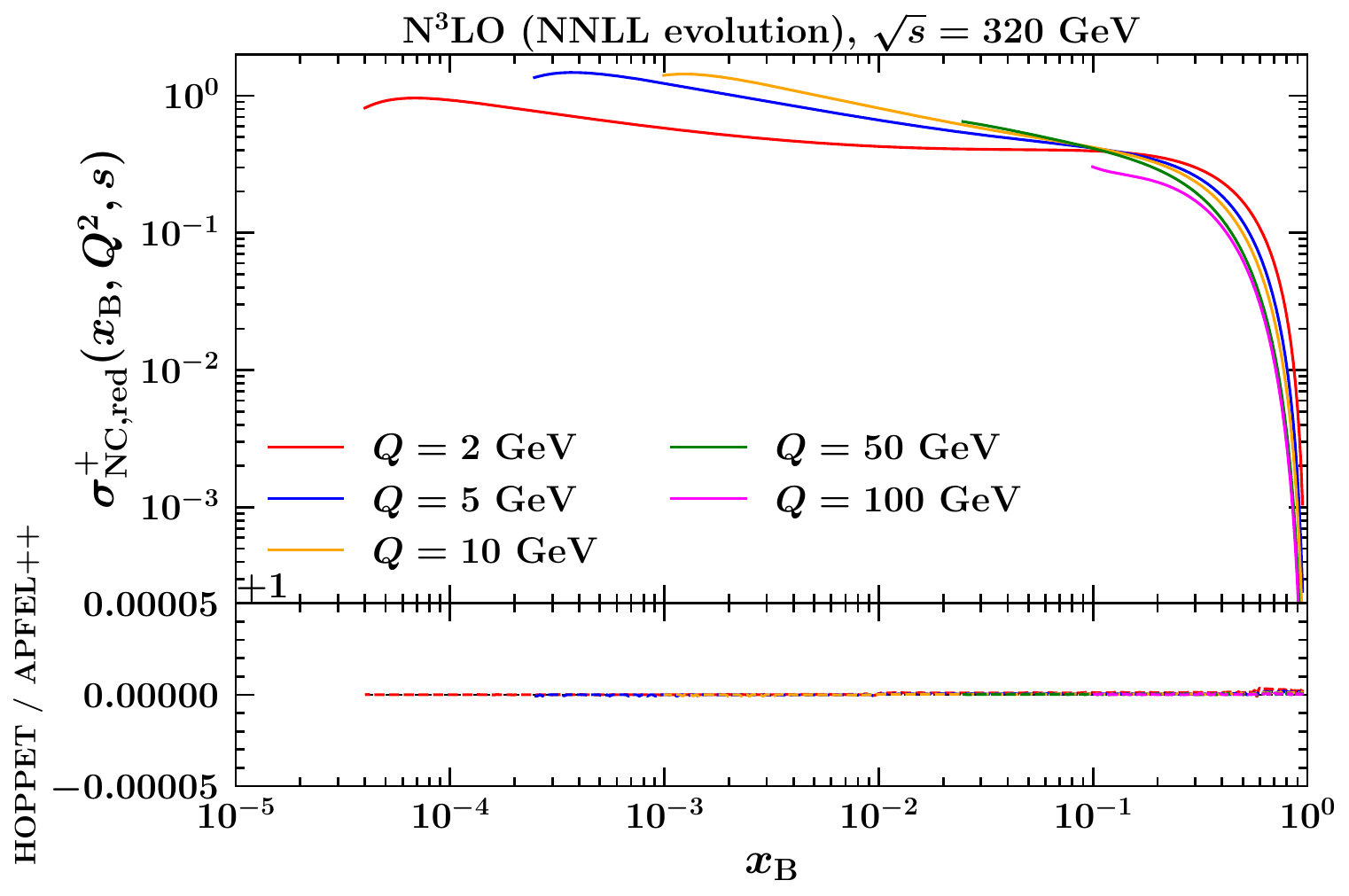}
  \centering\includegraphics[width=0.49\textwidth]{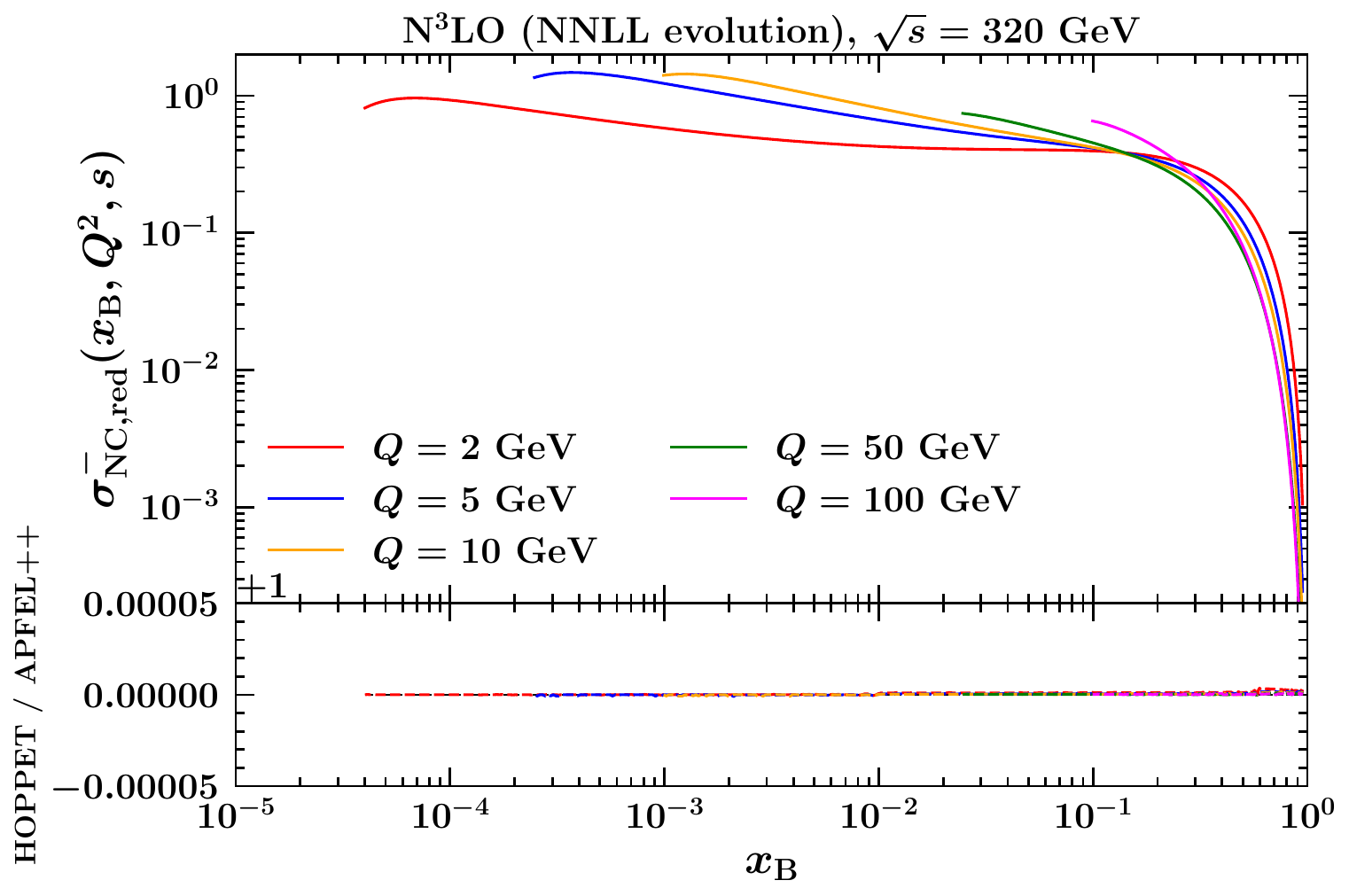}
  \centering\includegraphics[width=0.49\textwidth]{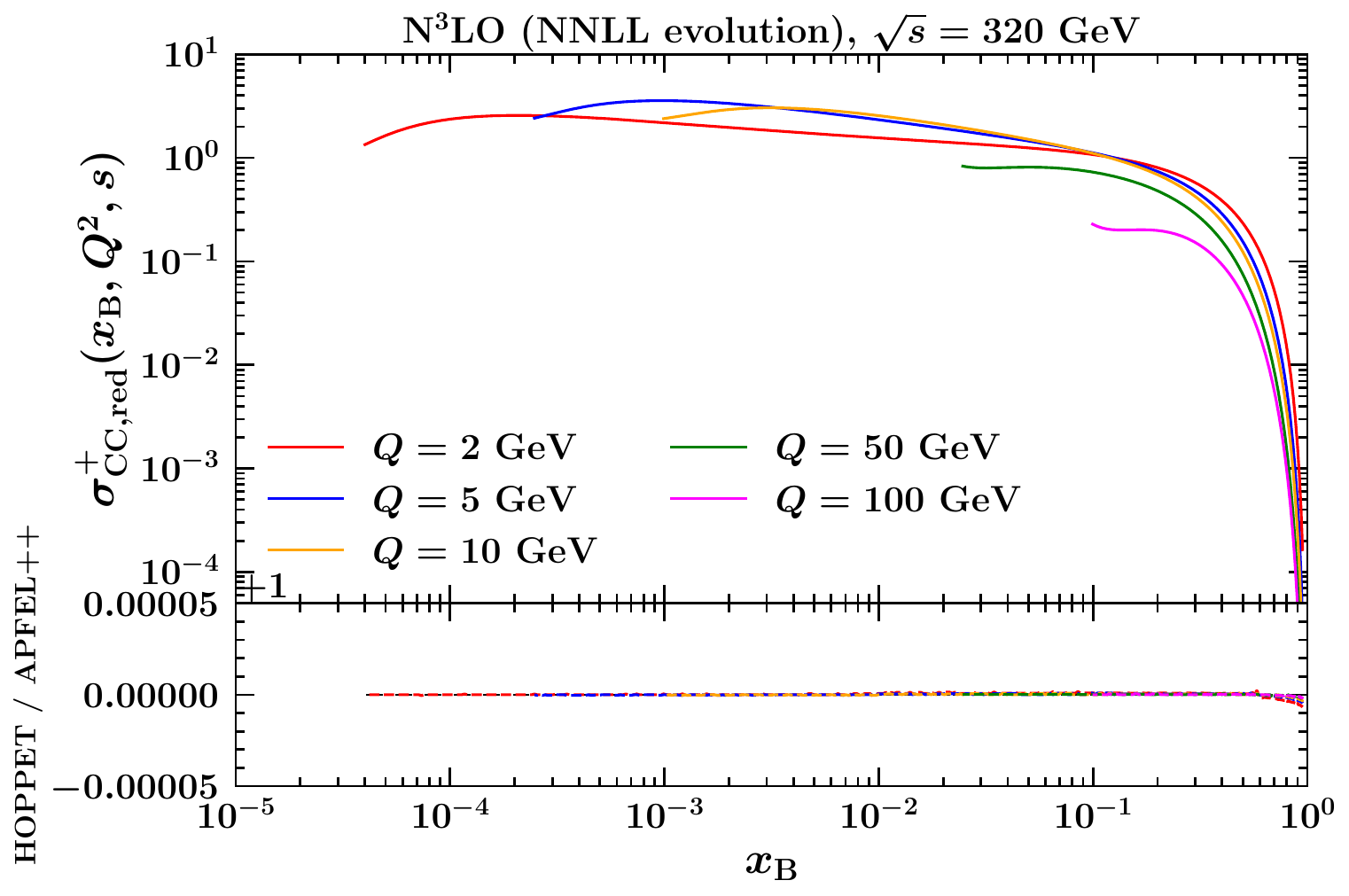}
  \centering\includegraphics[width=0.49\textwidth]{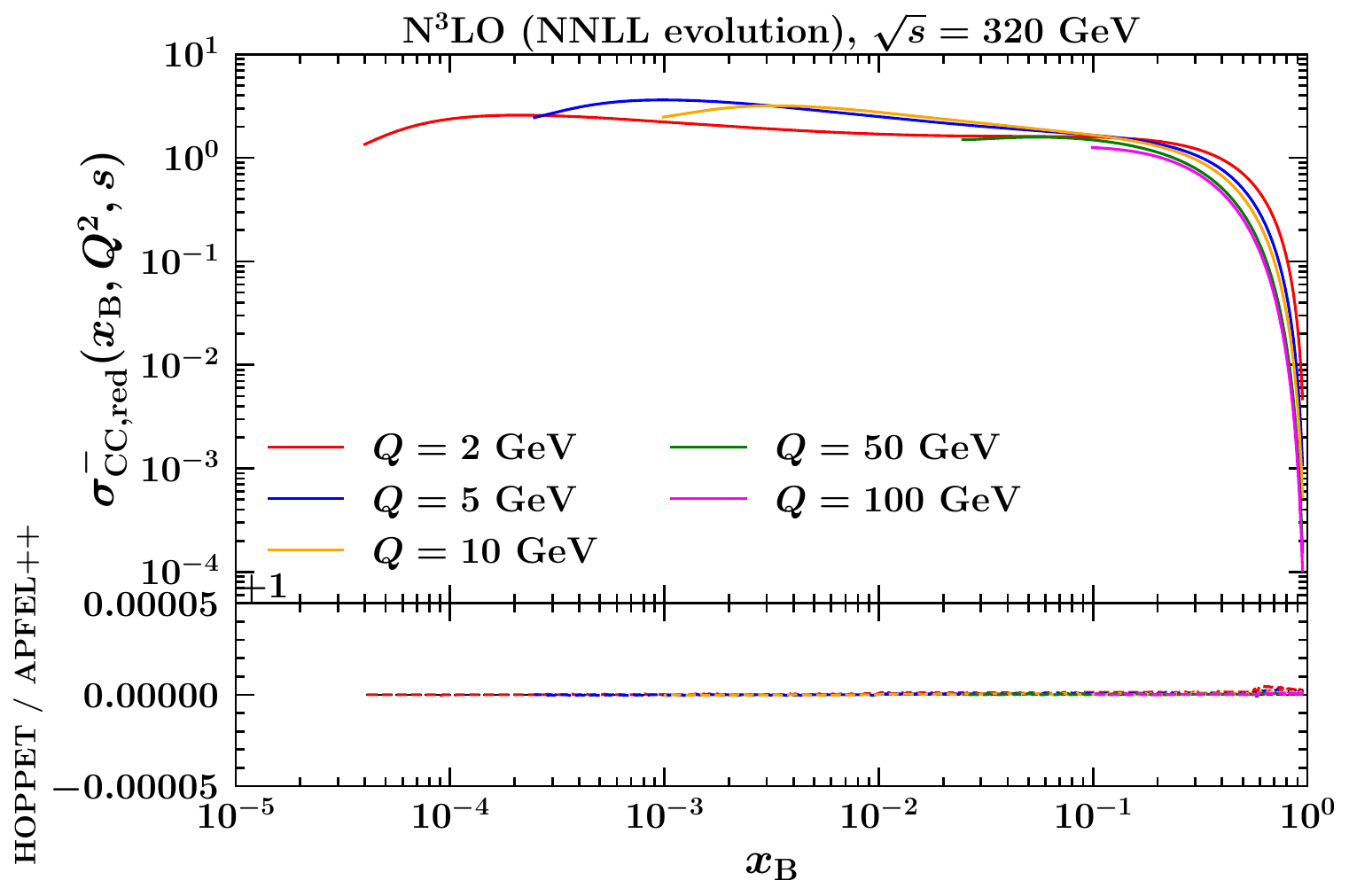}
  \caption{The reduced cross sections $\sigma_{\rm NC, red}^{+}$ (top
    left), $\sigma_{\rm NC, red}^{-}$ (top right),
    $\sigma_{\rm CC, red}^{+}$ (bottom left) and
    $\sigma_{\rm NC, red}^{-}$ (bottom right) at N$^3$LO as functions
    of $x_{\rm B}$ spanning between $10^{-5}$ and $0.9$ and for
    various values of the energy $Q$. The centre-of-mass energy is set
    to $\sqrt{s}=320$~GeV. The lower panels show the ratio between
    \apfel{} and \hoppet{}.}
  \label{fig:redxsec}
\end{figure}

As discussed in Sect.~\ref{sec:dis-sf}, it is also relevant to
consider the DIS reduced cross sections defined in
Eq.~\eqref{eq:redxsec}. As a matter of fact, the HERA collider has
delivered measurements for these observables~\cite{H1:2015ubc} that
are currently being employed in most of the modern PDF
determinations~\cite{Hou:2019efy, Bailey:2020ooq, NNPDF:2021njg}. In
Fig.~\ref{fig:redxsec}, we show N$^3$LO predictions for NC (top row)
and CC (bottom row) reduced cross sections relevant to $e^+p$ (left
column) and $e^-p$ (right column) collisions. The centre-of-mass
energy is set to $\sqrt{s}=320$~GeV, close to that of the latest runs
of HERA. A broad kinematic range in $x_{\rm B}$ and $Q$ is covered and
again we set $\mu_{\rm R}=\mu_{\rm F}=Q$. We notice that the curves,
presented as functions of $x_{\rm B}$ for different values of $Q$, are
limited in $x_{\rm B}$ by the physical requirement on the inelasticity
$y\leq 1$ (see Eq.~(\ref{eq:dis-variables})). As above, the lower
panel of each plot shows the ratio between predictions obtained with
\apfel{} and \hoppet{}. As expected from the results presented in
Fig.~\ref{fig:AllStructFuns}, the two codes agree well within
$10^{-5}$ relative accuracy over the full kinematic range also for the
reduced cross sections.

\begin{figure}[tb!]
  \centering\includegraphics[width=0.49\textwidth]{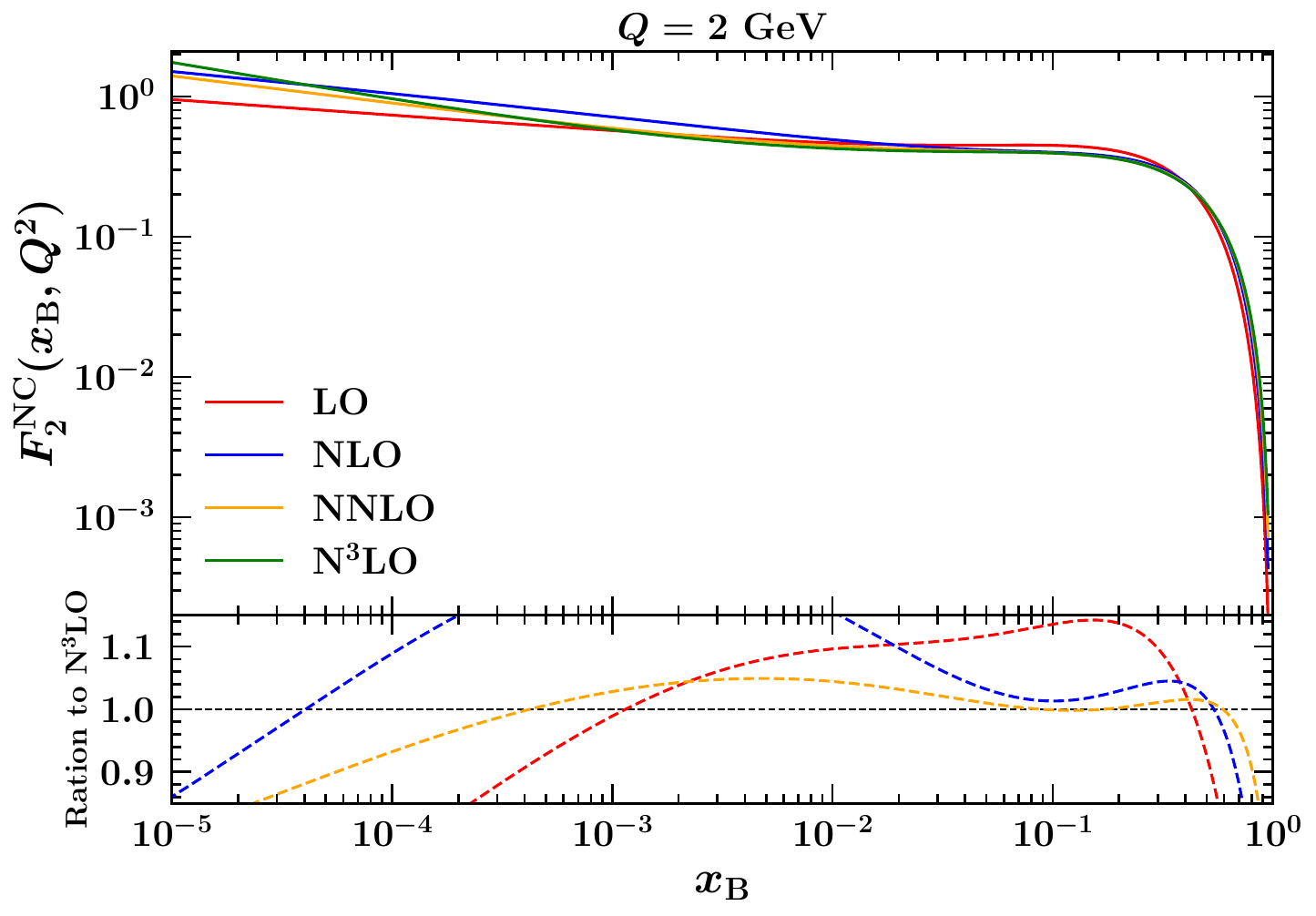}
  \centering\includegraphics[width=0.49\textwidth]{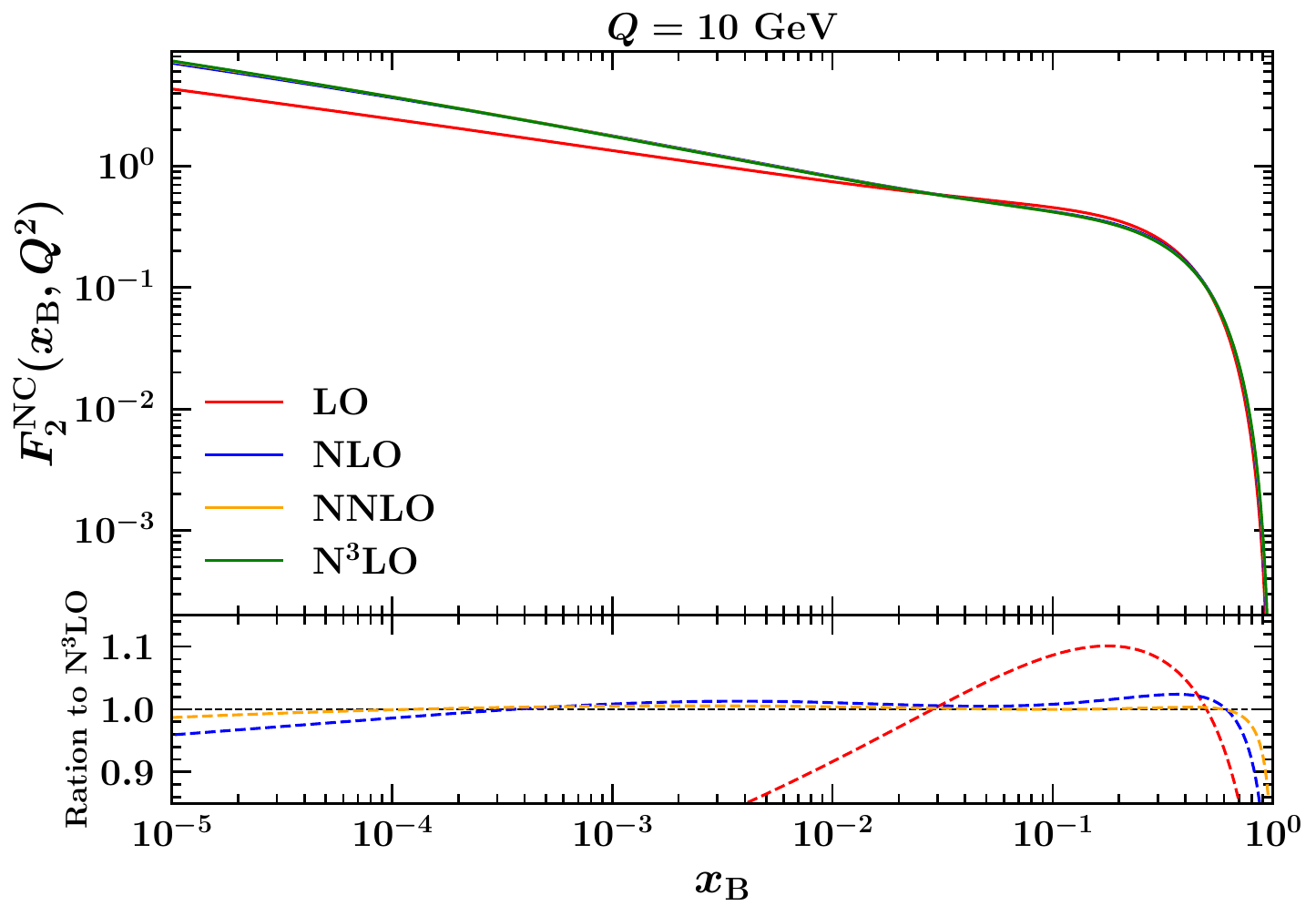}
  \caption{The structure function $F_2^{\rm NC}$ plotted as a function
    of $x_{\rm B}$ in the range $[10^{-5}:0.9]$ at $Q=2$~GeV (left)
    and $Q=10$~GeV (right). Each plot displays the curves at LO, NLO,
    NNLO, and N$^3$LO with the lower panel showing the ratio to
    N$^3$LO.}
  \label{fig:perturbativeconvergence}
\end{figure}

Having at our disposal four consecutive perturbative orders, it is
interesting to study how well the QCD perturbative series converges in
the case of inclusive DIS structure functions. In
Fig.~\ref{fig:perturbativeconvergence}, we display $F_2^{\rm NC}$ as a
function of $x_{\rm B}$ at $Q=2$~GeV (left) and $Q=10$~GeV (right)
computed with $\mu_{\rm R}=\mu_{\rm F}=Q$. Each plot shows this
structure function for all perturbative orders between LO and N$^3$LO,
with the lower panel giving the ratio to N$^3$LO. The pattern is
somewhat the expected one. At $Q=2$~GeV, due to the relatively large
value of $\alpha_s$, the convergence is slower with differences
between NNLO and N$^3$LO that can exceed 10\%, particularly at small
and large values of $x_{\rm B}$. At $Q=10$~GeV, instead, the
convergence is much faster with NNLO and N$^3$LO very close to each
other everywhere, except for very large values of $x_{\rm B}$.

\begin{figure}[tb!]
  \centering\includegraphics[width=0.49\textwidth]{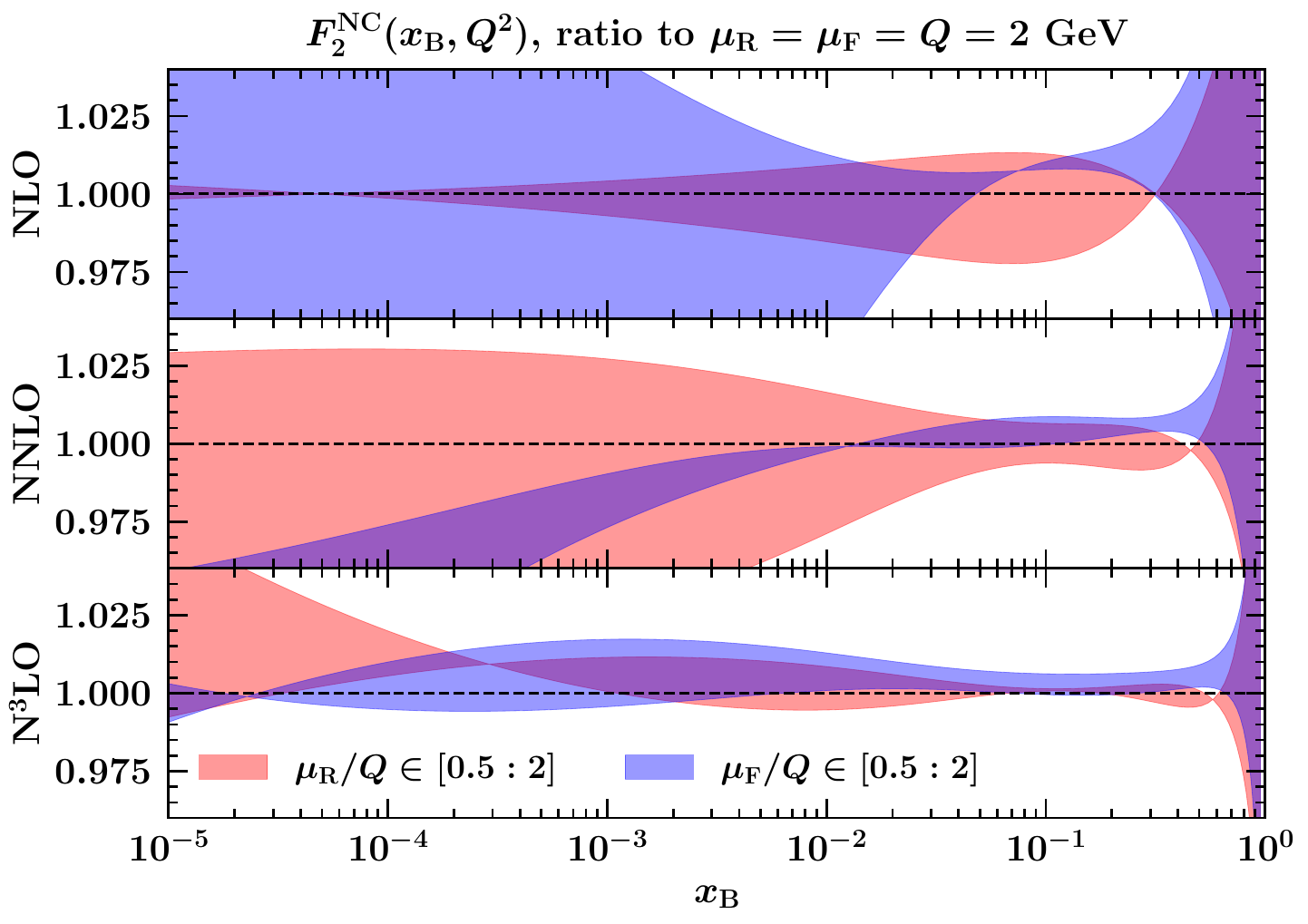}
  \centering\includegraphics[width=0.49\textwidth]{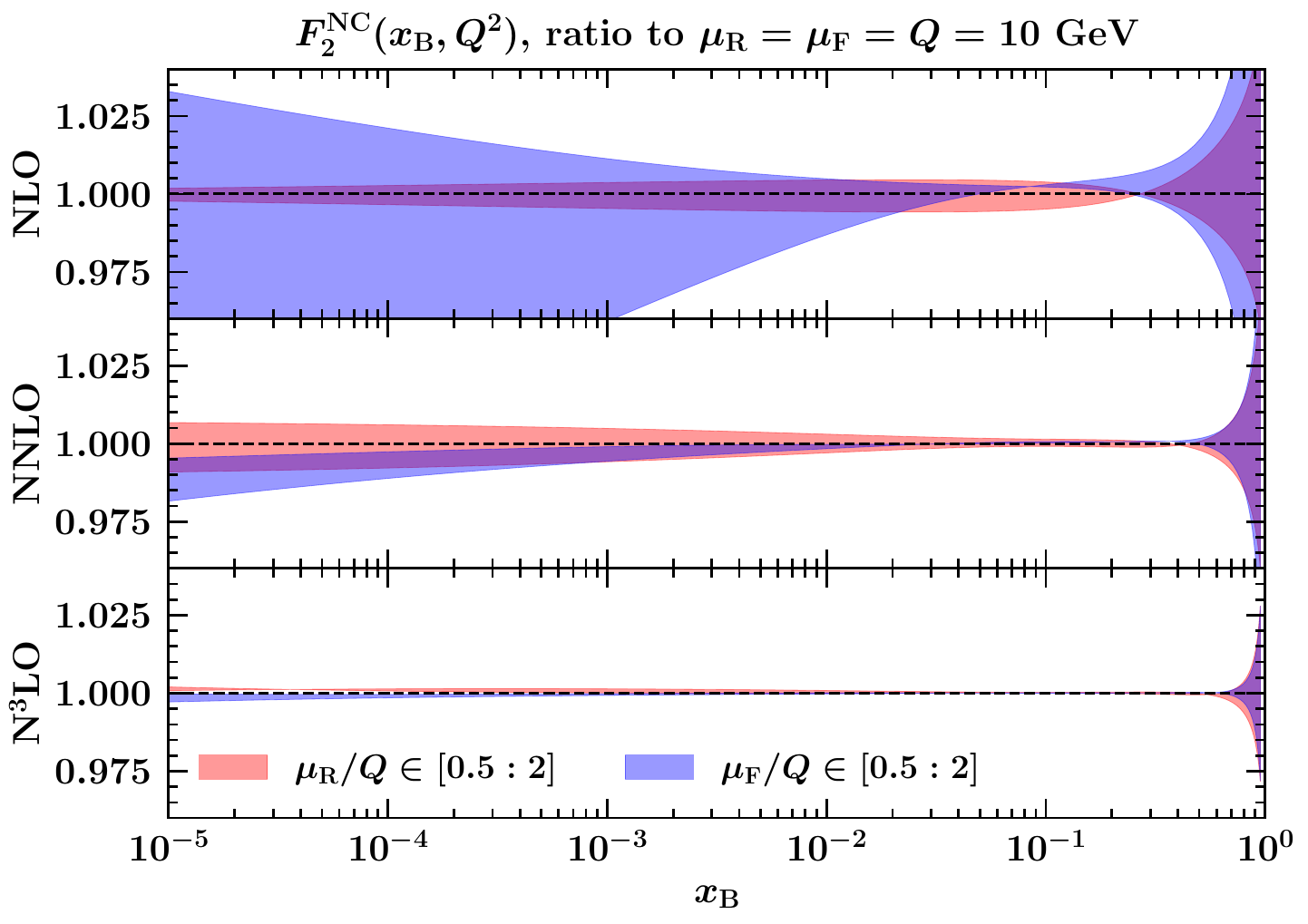}
  \caption{Relative scale variations on $F_2^{\rm NC}$ as functions of
    $x_{\rm B}$ for $Q=2$~GeV (left) and $Q=10$~GeV (right). The red
    (blue) bands correspond to variations of $\mu_{\rm R}$
    ($\mu_{\rm F}$) by a factor of $2$ up and down around
    $Q$. Variations are shown for NLO (upper panels), NNLO (central
    panels), and N$^3$LO (lower panels).}
  \label{fig:scalevariations}
\end{figure}

The perturbative convergence can also be studied by looking at how
renormalisation and factorisation scale variations behave. In
Sect.~\ref{sec:dis-sf}, we provided all relevant expressions to
perform these variations up to N$^3$LO accuracy. We point out that we
have checked that \apfel{} and \hoppet{} agree within the same level
of accuracy discussed above also when scales are varied. In
Fig.~\ref{fig:scalevariations}, we show the effect of varying the
renormalisation scale $\mu_{\rm R}$ (red bands) and the factorisation
scale $\mu_{\rm F}$ (blue bands) by a factor of $2$ up and down with
respect to $Q$ relative to the central-scale choice
$\mu_{\rm R}=\mu_{\rm F}=Q$. The left plot has been obtained with
$Q=2$~GeV while the right plot with $Q=10$~GeV. In each of them the
top panel shows variations at NLO, the central panel at NNLO, and the
bottom panel at N$^3$LO. We did not include the LO panel because at
this order inclusive DIS structure functions are independent of
$\mu_{\rm R}$ while $\mu_{\rm F}$ gives rise to very large
bands.\footnote{We notice that the bands shown in
  Fig.~\ref{fig:scalevariations} represent the area enclosed between
  the curves obtained with $\mu_{\rm R,F}/Q=1/2$ and
  $\mu_{\rm R,F}/Q=2$. It often happens that scale variations are not
  monotonic, such that the central-scale curve $\mu_{\rm R,F}/Q=1$
  falls outside the bands. In order to give a more reliable estimate
  of the perturbative uncertainty related to missing higher-order
  corrections, one should perform a scan between $\mu_{\rm R,F}/Q=1/2$
  and $\mu_{\rm R,F}/Q=2$ and quote the envelope as an
  uncertainty. However, here we do not mean to provide a realistic
  estimate of the scale uncertainties but we only want to study the
  perturbative convergence. Therefore, we limit to consider
  $\mu_{\rm R,F}/Q=1/2$ and $\mu_{\rm R,F}/Q=2$ only.} As expected,
scale-variation bands shrink significantly moving from NLO to N$^3$LO
at both scales. However, the reduction is much more pronounced at
$Q=10$~GeV than at $Q=2$~GeV, as a consequence of the decrease in
$\alpha_s$ value.

\section{Conclusion}
\label{sec:conclusion}

As N$^3$LO PDFs start to emerge, see \textit{e.g.}
Refs.~\cite{McGowan:2022nag,NNPDF:2024nan}, it will become
increasingly important to have reliable N$^3$LO predictions for
inclusive DIS cross sections available. Indeed, DIS measurements are
and will likely remain one of the main sources of experimental
information that enter modern determinations of PDFs. Moreover DIS is
one of the very few processes for which N$^3$LO corrections to the
partonic cross sections, albeit only in the quark massless limit, are
exactly known.

In this paper, we have benchmarked the implementation of the massless
DIS structure functions to N$^3$LO accuracy by comparing the
predictions provided by two widely used codes:
\apfel{}~\cite{Bertone:2013vaa,Bertone:2017gds} and
\hoppet{}~\cite{Salam:2008qg}. In this benchmark, we considered both
NC and CC structure functions relevant to the computation of DIS cross
sections respectively characterised by the exchange of a neutral
($\gamma/Z$) and a charged ($W^{\pm}$) virtual vector boson. The
numerical setup closely follows that of
Ref.~\cite{Giele:2002hx}. Specifically, we used a realistic set of
initial-scale PDFs that were evolved to the relevant scales before
convoluting them with the appropriate DIS coefficient functions. This
workflow was independently implemented in both \apfel{} and \hoppet{}
before comparing the respective predictions. On top of the single
structure functions, we also compared reduced cross sections as those
delivered by the HERA experiments~\cite{H1:2015ubc}.

We found a relative agreement between \apfel{} and \hoppet{} at the
$10^{-5}$ level or better over a very wide kinematic range in
$Q\in[2:100]$~GeV and $x_{\rm B}\in[10^{-5}:0.9]$ for all structure
functions and reduced cross sections. Additionally, we also investigated
the perturbative convergence by comparing predictions for
$F_{2}^{\rm NC}$ at all available perturbative orders, \textit{i.e.}
from LO to N$^3$LO, and by estimating the effect of renormalisation
and factorisation scale variations. We found the expected pattern
according to which $F_{2}^{\rm NC}$ exhibits a good perturbative
convergence, especially for large values of $Q$. \apfel{} and
\hoppet{} were benchmarked also against scale variations.

The benchmark carried out in this paper, thanks to its high accuracy
level, provides a solid reference for any future implementation of the
DIS coefficient functions up to N$^3$LO perturbative accuracy. In
order to make our results fully reproducible and facilitate the
comparison to future implementations, we made the code used for this
benchmark available at:
\begin{center}
  {\tt
    \href{https://github.com/alexanderkarlberg/n3lo-structure-function-benchmarks}{https://github.com/alexanderkarlberg/n3lo-structure-function-benchmarks}},
\end{center}
where we also provide a short documentation and the suite of {\tt
  Matplotlib} scripts that we used to produce the plots shown in this
paper.

\section*{Acknowledgments}

We are grateful to Gavin Salam for many fruitful discussions and a
critical reading of the paper. A.~K. is also grateful to Frédéric
Dreyer for the initial work on the structure functions in
\hoppet{}. V.~B. is supported by the European Union Horizon 2020
research and innovation program under grant agreement Num. 824093
(STRONG-2020).

\appendix

\section{Comments on the $y\to 1$ behaviour of the coefficient functions}
\label{app:exact-vs-param}

In the course of this benchmark, we have encountered some minor
differences between the exact N$^3$LO coefficient functions and the
parameterisations used here and given in
Refs.~\cite{Moch:2004xu,Vermaseren:2005qc,Vogt:2006bt,Moch:2008fj}. The
differences arise for $y \gtrsim 0.9$ in the regular part of the
non-singlet coefficient functions.\footnote{Specifically, in the
  routines {\tt CLNP3A}, {\tt C2NP3A}, and {\tt C3NM3A} of {\tt
    xclns3p.f}, {\tt xc2ns3p.f}, and {\tt xc3ns3p.f}, respectively. }
Although the difference is phenomenologically negligible, it does
have a small impact in precision studies like the benchmark presented
here. The largest relative difference is found in the $c^{(3)}_{2,q}$
coefficient function, on which we focus our attention.

In general, a DIS coefficient function $C$ receives three different
contributions:
\begin{itemize}
\item The \emph{singular} piece $[C_{\rm sing}(y)]_+$, which is a
  combination of terms of the kind
  $\left[\frac{\ln^i(1-y)}{1-y}\right]_{+}$. The $+$-prescription is
  defined as:
\begin{equation}
  \int_0^1 dy [C_{\rm sing}(y)]_+ f(y) = \int_0^1 dy \, C_{\rm sing}(y) [f(y) - f(1)]\,,
\end{equation} 
and has the effect of regularising the otherwise non-integrable
singularities at $y=1$.

\item The \emph{regular} piece $C_{\rm reg}(y)$, which in general can
  be very complicated but in the $y\rightarrow 1$ limit develops
  integrable singular terms of the kind $\ln^i(1-y)$.

\item The \emph{local} piece $C_{\rm loc}\delta(1-y)$, where
  $C_{\rm loc}$ is a numerical constant.
\end{itemize}

Although the singular piece $[C_{\rm sing}(y)]_+$ dominates the
coefficient function for $y\rightarrow 1$, this is not the case when
it is convoluted with a parton distribution as in
Eq.~\eqref{eq:conv-structf}, since the $+$-prescription effectively
generates a factor of $1-y$. In the limit $y\rightarrow 1$, this can
be seen schematically as follows:
\begin{align}
  \label{eq:app-convolution}
  C\otimes f & =  \int_{x_{\rm B}}^1 \frac{dy}{y}\left\{C_{\mathrm{reg}}(y) +
               \left[C_{\mathrm{sing}}(y)\right]_++C_{\rm loc}\delta(1-y)\right\}f\left(\frac{x_{\rm B}}{y}\right) \notag \\
             & \simeq \int_{x_{\rm B}}^1 dy \left\{C_{\mathrm{reg}}(y)f(x_{\rm B}) + (1-y)C_{\mathrm{sing}}(y)\left[f(x_{\rm B})+x_{\rm B} f'(x_{\rm B})\right] + \dots \right\}\,,
\end{align}
where in the second line we have expanded $f(x_{\rm B}/y)/y$ around
$y=1$ and neglected the local piece as well as the additional terms
proportional to $\ln^{i+1}(1-x_{\rm B})\delta(1-y)$ generated by the
$+$-prescription. Therefore, in order to achieve accurate results when
parameterising $C$, it is necessary to correctly account for the
large-$y$ behaviour of $C_{\rm reg}$.

\begin{figure}[tb!]
  \centering\includegraphics[width=0.49\textwidth, page=2]{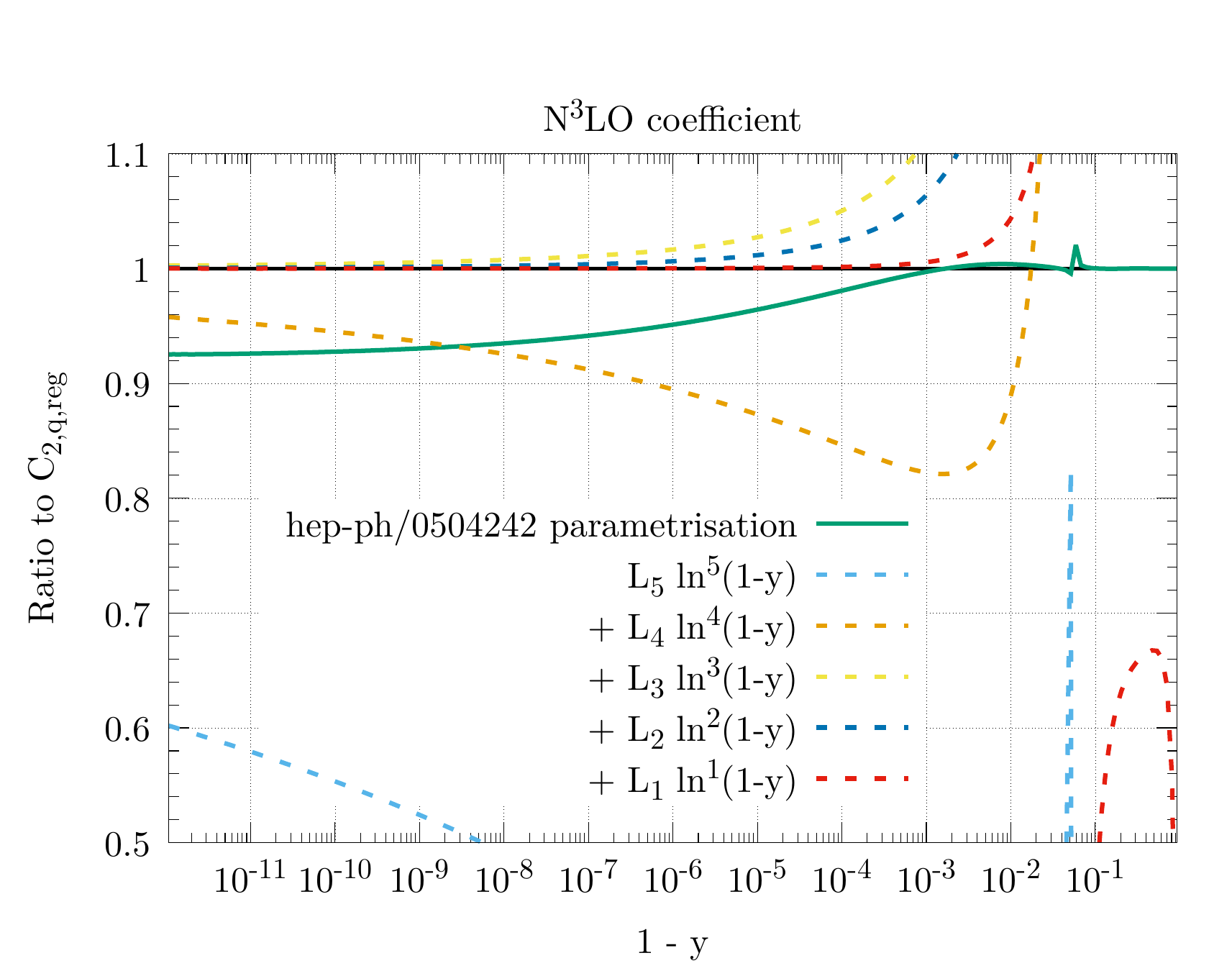}
  \centering\includegraphics[width=0.49\textwidth, page=3]{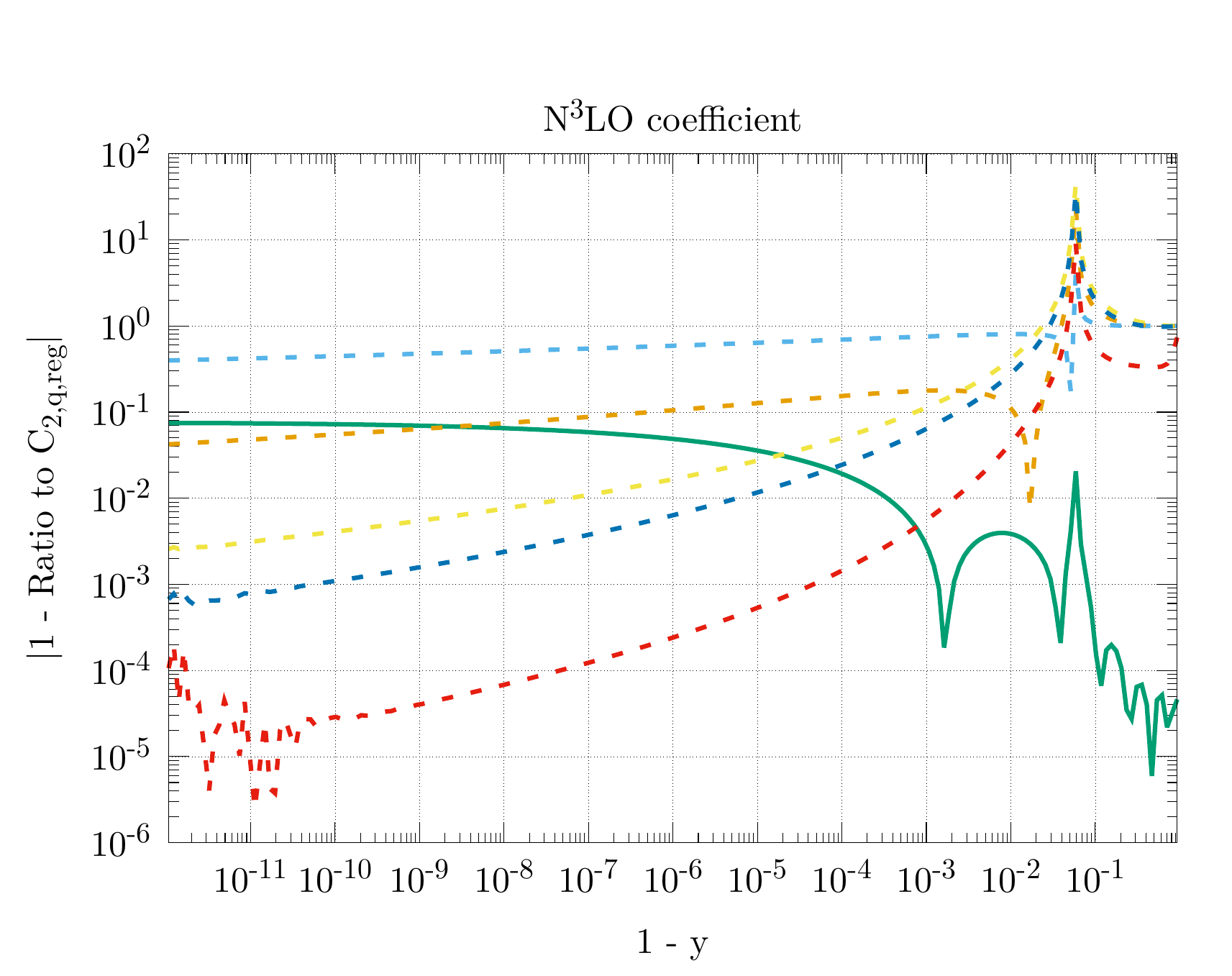}
  \caption{Left: The regular $\mathcal{O}(\as^3)$ coefficient function
    $c^{(3)}_{2,\mathrm{q, reg}}$ as a function of $1-y$ plotted as a
    ratio to the exact expression. In green we show the
    parametrisation of Ref.~\cite{Vermaseren:2005qc}. The dashed lines
    show the large-$y$ expansion including progressively more
    terms. Right: The same plot but now showing $1$ minus the ratio on
    a log-scale to highlight the relative agreement.}
  \label{fig:exactvsparam}
\end{figure}

The solid green curve on the l.h.s. of Fig.~\ref{fig:exactvsparam}
shows the ratio between the regular part of the parametrisation for
$c_{2,q}^{(3)}$ given in Eq.~(4.11) of Ref.~\cite{Vermaseren:2005qc}
and its exact counterpart as a function of $1-y$. As can be seen, the
ratio increases as $1-y$ approaches zero and, in the range shown on
the plot, it reaches $8\%$.

In order to investigate this difference, we considered the large-$y$
limit of the regular part of all $\mathcal{O}(\alpha_s^3)$ non-singlet
coefficient functions, which in this region admit the following
expansion:
\begin{align}
  c_{k,q,{\rm reg}}^{(3)}(y) \simeq \sum_{i=1}^5 L_{i}^{(k)} \ln^i(1-y)\,,
\end{align}
with $k=2,3,L$. The coefficients $L_i$ can be found in
Refs.~\cite{Vermaseren:2005qc,Moch:2008fj}. Since these coefficients
are non-trivial to derive, we recomputed them finding two minor typos
in the expressions for $c^{(3)}_{L,q,{\rm reg}}$ reported in
Ref.~\cite{Vermaseren:2005qc}. Specifically, for the coefficients of
$c^{(3)}_{2,q,{\rm reg}}$ we find:
\begin{eqnarray}
\label{c23L15}
  L_5^{(2)}\!\!\! & = \! &
          - \: 8\: \* \cft\,,
\\[0.5mm]
\label{c23L14}
  L_4^{(2)} \!\!\! & = \! &
            {220 \over 9} \: \* \ca \* \cfs
       \: + \: 92\: \* \cft
       \: - \: {40 \over 9} \: \* \cfs\, \* \nf\,,
\\[1mm]
\label{c23L13}
  L_3^{(2)}\!\!\! & = \! &
       - \: {484 \over 27}\: \* \cas \* \cf
       - \: \ca \* \cfs \*  \:\Bigg[
            {10976 \over 27}
          - 64\, \* \z2
         \Bigg]
     \: - \: \cft  \* \: [
            38
          - 32\, \* \z2
          ]
\nn \\[1mm] & & \mbox{\hspn}
     \: + \: {176 \over 27}\: \* \ca \* \cf\, \* \nf  
     \: + \: {1832 \over 27}\: \* \cfs\, \* \nf
     \: - \: {16 \over 27}\: \* \cf\, \* \nfs\,,
\\[2mm]
\label{c23L12}
  L_2^{(2)}\!\!\! & = \! &
          \cas \* \cf \* \:\Bigg[
            {11408 \over 27}
          - {266 \over 3}\: \* \z2
          - 32\, \* \z3
         \Bigg]
     \: + \: \ca \* \cfs \* \:\Bigg[ 
            {11501 \over 9}
          - 292\, \* \z2
          - 160\, \* \z3
          \Bigg]
\nn\\[1mm] & & \mbox{\hspn}
     - \: \cft \* \:\Bigg[
            {1199 \over 3}
          + 688\, \* \z2
          + 48\, \* \z3
          \Bigg]
     \: - \: \ca \* \cf\, \* \nf \* \:\Bigg[ 
            {3694 \over 27}
          - {64 \over 3}\: \* \z2
          \Bigg]
\nn\\[1mm] & & \mbox{\hspn}
     - \: \cfs\, \* \nf \* \:\Bigg[
            {2006 \over 9}
          - {16 \over 3}\, \* \z2
          \Bigg]
     \: + \: {296 \over 27}\: \* \cf\, \* \nfs\,,
\\[2mm]
\label{c23L11}
  L_1^{(2)}\!\!\! & = \! &
        - \: \cas \* \cf \* \:\Bigg[
            {215866 \over 81}
          - 824\: \* \z2
          - {1696 \over 3}\, \* \z3
          + {304 \over 5}\: \* \zss
         \Bigg]
     \: + \: \ca \* \cfs \* \:\Bigg[
            {126559 \over 162}
\nn\\[1mm] & & 
          + 872\: \* \z2
          + 792\: \* \z3
          - {1916 \over 5}\: \* \zss
          \Bigg]
     \: + \: \cft  \*  \:\Bigg[
            {157 \over 6}
          + {1268 \over 3}\: \* \z2
\nn\\[1mm] & & 
          - 272\, \* \z3
          + 488\: \* \zss
          \Bigg]
     \: + \: \ca \* \cf\, \* \nf \* \:\Bigg[
            {64580 \over 81}
          - {1292 \over 9}\, \* \z2
          - {304 \over 3}\, \* \z3
          \Bigg]
\nn\\[1mm] & & \mbox{\hspn}
     \: - \: \cfs\, \* \nf \* \:\Bigg[
            {4445 \over 81}
          + 208\: \* \z2
          - {208 \over 3}\: \* \z3
          \Bigg]
     \: - \: \cf\, \* \nfs \* \:\Bigg[
            {4432 \over 81}
          - {32 \over 9}\: \* \z2
          \Bigg]\,,
\end{eqnarray}
which agree with the results of
Ref.~\cite{Moch:2008fj}. Similarly, for the coefficients of
$c^{(3)}_{3,q, {\rm reg}}$ we have:
\begin{eqnarray}
\label{c33L15}
 L_5^{(3)} \!\!\! & = \! &
          - \: 8\: \* \cft\,,
\\[0.5mm]
\label{c33L14}
  L_4^{(3)}\!\!\! & = \! &
            {220 \over 9} \: \* \ca \* \cfs
       \: + \: 84\: \* \cft
       \: - \: {40 \over 9} \: \* \cfs\, \* \nf\,,
\\[1mm]
\label{c33L13}
  L_3^{(3)}\!\!\! & = \! &
       - \: {484 \over 27}\: \* \cas \* \cf
       - \: \ca \* \cfs \*  \:\Bigg[
            {9056 \over 27}
          - 32\, \* \z2
         \Bigg]
     \: - \: \cft  \* \: [
            110
          - 96\, \* \z2
          ]
\nn \\[1mm] & & \mbox{\hspn}
     \: + \: {176 \over 27}\: \* \ca \* \cf\, \* \nf  
     \: + \: {1640 \over 27}\: \* \cfs\, \* \nf
     \: - \: {16 \over 27}\: \* \cf\, \* \nfs\,,
\\[2mm]
\label{c33L12}
  L_2^{(3)}\!\!\! & = \! &
          \cas \* \cf \* \:\Bigg[
            {7580 \over 27}
          - {98 \over 3}\: \* \z2
         \Bigg]
     \: + \: \ca \* \cfs \* \:\Bigg[ 
            {12031 \over 9}
          - 372\, \* \z2
          - 240\, \* \z3
          \Bigg]
\nn\\[1mm] & & \mbox{\hspn}
     - \: \cft \* \:\Bigg[
            {1097 \over 3}
          + 656\, \* \z2
          + 16\, \* \z3
          \Bigg]
     \: - \: \ca \* \cf\, \* \nf \* \:\Bigg[ 
            {2734 \over 27}
          - {16 \over 3}\: \* \z2
          \Bigg]
\nn\\[1mm] & & \mbox{\hspn}
     - \: \cfs\, \* \nf \* \:\Bigg[
            {2098 \over 9}
          - {112 \over 3}\, \* \z2
          \Bigg]
     \: + \: {248 \over 27}\: \* \cf\, \* \nfs\,,
\\[2mm]
\label{c33L11}
  L_1^{(3)}\!\!\! & = \! &
        - \: \cas \* \cf \* \:\Bigg[
            {138598 \over 81}
          - {4408 \over 9}\: \* \z2
          - 272\, \* \z3
          + {176 \over 5}\: \* \zss
         \Bigg]
     \: - \: \ca \* \cfs \* \:\Bigg[
            {69833 \over 162}
\nn\\[1mm] & & 
          - \: {12568 \over 9}\: \* \z2
          - {1904 \over 3}\: \* \z3
          + {764 \over 5}\: \* \zss
          \Bigg]
     \: + \: \cft  \*  \:\Bigg[
            {1741 \over 6}
          + {1220 \over 3}\: \* \z2
\nn\\[1mm] & & 
          + 480\, \* \z3
          - {376 \over 5}\: \* \zss
          \Bigg]
     \: + \: \ca \* \cf\, \* \nf \* \:\Bigg[
            {45260 \over 81}
          - 108\, \* \z2
          - \: 16\, \* \z3
          \Bigg]
\nn\\[1mm] & & \mbox{\hspn}
     \: + \: \cfs\, \* \nf \* \:\Bigg[
            {9763 \over 81}
          - {2224 \over 9}\: \* \z2
          - {112 \over 3}\: \* \z3
          \Bigg]
     \: - \: \cf\, \* \nfs \* \:\Bigg[
            {3520 \over 81}
          - {32 \over 9}\: \* \z2
          \Bigg]\,,
\end{eqnarray}
which agree with the results in Ref.~\cite{Moch:2008fj}. Finally, for
the coefficients of $c^{(3)}_{L,q, {\rm reg}}$ we find:
\begin{eqnarray}
\label{eq:clql14}
  L_4^{(L)}\! & = \! &
    8\: C_F^{\,3} \:\: , \\[2mm]
\label{eq:clql13}
  L_3^{(L)}\! & = \! &
  C_A C_F^{\,2} \left[ - \frac{640}{9} + 32\:\z2 \right] \: + \:
  C_F^{\,3} \Big[ 72 - 64\:\z2 \Big] \: {\color{blue}+ \: \frac{64}{9}\: C_F^2 \nf} 
  \:\: , \\[2mm]
\label{eq:clql12}
  L_2^{(L)}\! & = \! &
  C_A^{\,2} C_F \,\left[ \frac{1276}{9} - 56\:\z2 - 32\:\z3 \right]
     \: + \:
  C_A  C_F^{\,2} \,\left[ -\frac{530}{9} + 80\:\z2 + 80\:\z3 \right]
     \nn \\[0.5mm] & & \mbox{} + \:
  C_F^{\,3} \,\Big[ -34 - 32\:\z2 -32\:\z3 \Big]
     \: + \:
  C_A C_F \nf \,\left[ -\frac{320}{9} + 16\:\z2 \right]
     \nn \\[0.5mm] & & \mbox{} + \:
  C_F^{\,2} \nf \,\left[ \frac{92}{9} - 32\:\z2 \right]
     \: + \:  \frac{16}{9}\: C_F \n2f  \:\: , \\[2mm]
\label{eq:clql11}
  L_1^{(L)}\! & = \! &
  C_A^{\,2} C_F \,\left[ - \frac{25756}{27} + \frac{3008}{9}\:\z2 
     + \frac{880}{3}\:\z3 - \frac{128}{5}\:\zs\right] 
  \nn \\[0.5mm] & & \mbox{} + \:
  C_A C_F^{\,2} \,\left[ \frac{32732}{27} - \frac{4720}{9}\:\z2 
     + \frac{472}{3}\:\z3 - \frac{1152}{5}\:\zs \right] \: + \: 
  C_F^{\,3} \,\Bigg[ - 264 
     \nn \\[0.5mm] & & \left. \mbox{} 
     + 16\:\z2 - 752\:\z3 {\color{blue}+ \frac{2816}{5}\:\zs} \right] \: + \:
  C_A C_F \nf \,\left[ \frac{6640}{27} - \frac{320}{9}\:\z2 
     - \frac{256}{3}\:\z3 \right]
     \nn \\[0.5mm] & & \mbox{} + \:
  C_F^{\,2} \nf \,\left[ -\frac{4736}{27} + \frac{352}{9}\:\z2 
     + \frac{320}{3}\:\z3 \right]
     \: - \:  \frac{304}{27}\: C_F \n2f \:\: , \\[2mm]
\end{eqnarray}
which also agree with the expressions given in
Ref.~\cite{Vermaseren:2005qc}, except for the two terms highlighted in
blue.

On the l.h.s. plot of Fig.~\ref{fig:exactvsparam}, we also show the
large-$y$ expansion of Eqs.~\eqref{c23L15}--\eqref{c23L11} retaining
progressively more terms. It can be seen that the agreement between
the exact coefficient function and the large-$y$ expansion improves as
$y$ increases, as expected. We conclude that the parameterisation for
the regular part of $c_{2,q}^{(3)}$ reported in Eq.~(4.11) of
Ref.~\cite{Vermaseren:2005qc} does not fully account for its large-$y$
behaviour, but we also stress that the phenomenological impact is
negligible.

On the r.h.s. of Fig.~\ref{fig:exactvsparam}, we also show $1$ minus
the ratio to the exact expression to highlight the \textit{relative}
agreement between the various curves. From this plot, we notice an
apparent degradation of the overall precision of the exact expression
as provided in Ref.~\cite{Vermaseren:2005qc} as $y$ approaches one
(see the oscillations of the red curve).\footnote{The Fortran
  implementation of this expression relies on a weight-5 extension of
  the {\tt hplog} package~\cite{Gehrmann:2001pz} for the evaluation of
  the harmonic polylogarithms. We have explicitly checked that the
  decrease in precision is not due to this evaluation, as it persists
  also when using the {\tt HPOLY} program~\cite{Ablinger:2018sat}.} In
\hoppet{}, where the exact expressions have been implemented, this
causes some issues as the numerical convolution requires the
evaluation of the coefficient functions at values of $y$ that in
double precision are indistinguishable from $1$. For this reason, we
switch to the large-$y$ expressions close to $y=1$.

As stated above, the difference between using exact and parametrised
coefficient functions is phenomenologically negligible at the level of
the structure functions. Indeed, if we reproduce the benchmark
tables~\ref{tab:N2LO-Q2}--\ref{tab:N3LO-Q100} using the exact N$^3$LO
expressions in \hoppet{} (but keeping the parametrisations at NNLO),
they typically differ from the tables obtained with the parameterised
expressions only in the last (fifth) digit, and often not at all. Our
benchmark program, \codelink{StructureFunctionsJoint.cc}, can be
modified to use the exact expressions in \hoppet{} by setting the flag
{\tt param\_coefs} to {\tt false}. Finally, we provide a Fortran file,
\auxlink{c\_ns\_reg\_large\_x.f}, with the large-$y$ expressions given
above.

\section{Benchmark tables}
\label{app:bench_tables}

In this appendix, we collect benchmark tables for all of the inclusive
DIS structure functions. Results are presented for both NC and CC
channels at NLO, NNLO, and N$^3$LO accuracy, for $Q=2, 50, 100$~GeV,
and at values of $x_{\rm B}$ ranging from $10^{-5}$ to $0.9$. Details
on the numerical setup can be found in Sect.~\ref{sec:setup}. The
numbers reported in Tabs.~\ref{tab:N1LO-Q2}--\ref{tab:N3LO-Q100} agree
between \apfel{} and \hoppet{} within the digits shown. The code that
produces the tables can be found in
\codelink{StructureFunctionsJoint.cc}.

\begin{table}[h]
\begin{adjustbox}{width=\textwidth}
\begin{tabular}{|c||c|c|c|c|c|c|c|c|c|}
\hline
$x_{\rm B}$ & $F_1^{\rm NC}$ & $F_2^{\rm NC}$ & $F_3^{\rm NC}$ & $F_1^{W^+}$ & $F_2^{W^+}$ & $F_3^{W^+}$ & $F_1^{W^-}$ & $F_2^{W^-}$ & $F_3^{W^-}$ \\
\hline
$ 1.0^{-5}$ & $ 4.9966^{+4}$ & $ 1.5085^{+0}$ & $ 3.4458^{-2}$ & $ 9.6690^{+4}$ & $ 9.6717^{+4}$ & $ 2.8589^{+0}$ & $ 2.8595^{+0}$ & $ 2.1997^{+4}$ & $-2.1814^{+4}$ \\
$ 1.0^{-4}$ & $ 3.5370^{+3}$ & $ 1.0505^{+0}$ & $ 1.7875^{-2}$ & $ 6.8928^{+3}$ & $ 6.9067^{+3}$ & $ 2.0028^{+0}$ & $ 2.0058^{+0}$ & $ 1.7751^{+3}$ & $-1.6803^{+3}$ \\
$ 1.0^{-3}$ & $ 2.4761^{+2}$ & $ 7.1563^{-1}$ & $ 8.7463^{-3}$ & $ 4.8255^{+2}$ & $ 4.8939^{+2}$ & $ 1.3664^{+0}$ & $ 1.3811^{+0}$ & $ 1.5419^{+2}$ & $-1.0784^{+2}$ \\
$ 1.0^{-2}$ & $ 1.8294^{+1}$ & $ 4.9277^{-1}$ & $ 3.9543^{-3}$ & $ 3.3637^{+1}$ & $ 3.6713^{+1}$ & $ 9.0065^{-1}$ & $ 9.6775^{-1}$ & $ 1.7954^{+1}$ & $ 2.9668^{+0}$ \\
$ 1.0^{-1}$ & $ 1.7795^{+0}$ & $ 4.0072^{-1}$ & $ 1.4381^{-3}$ & $ 2.4287^{+0}$ & $ 3.5994^{+0}$ & $ 5.5218^{-1}$ & $ 8.0596^{-1}$ & $ 3.1699^{+0}$ & $ 4.3670^{+0}$ \\
$ 3.0^{-1}$ & $ 5.0019^{-1}$ & $ 3.1315^{-1}$ & $ 5.3436^{-4}$ & $ 4.8044^{-1}$ & $ 1.0247^{+0}$ & $ 3.0027^{-1}$ & $ 6.4138^{-1}$ & $ 8.7182^{-1}$ & $ 1.8494^{+0}$ \\
$ 5.0^{-1}$ & $ 1.6809^{-1}$ & $ 1.7140^{-1}$ & $ 1.8517^{-4}$ & $ 1.1350^{-1}$ & $ 3.5189^{-1}$ & $ 1.1539^{-1}$ & $ 3.5888^{-1}$ & $ 2.2211^{-1}$ & $ 6.8705^{-1}$ \\
$ 7.0^{-1}$ & $ 3.7649^{-2}$ & $ 5.3128^{-2}$ & $ 4.0874^{-5}$ & $ 1.5406^{-2}$ & $ 8.0950^{-2}$ & $ 2.1703^{-2}$ & $ 1.1424^{-1}$ & $ 3.0736^{-2}$ & $ 1.6129^{-1}$ \\
$ 9.0^{-1}$ & $ 1.6475^{-3}$ & $ 2.9702^{-3}$ & $ 1.7405^{-6}$ & $ 2.2898^{-4}$ & $ 3.6497^{-3}$ & $ 4.1266^{-4}$ & $ 6.5798^{-3}$ & $ 4.5806^{-4}$ & $ 7.2985^{-3}$ \\
\hline
\end{tabular}
\end{adjustbox}\caption{NLO stucture functions with NLL evolution at $Q = 2$ GeV.}
\label{tab:N1LO-Q2}
\end{table}

\begin{table}[h]
\begin{adjustbox}{width=\textwidth}
\begin{tabular}{|c||c|c|c|c|c|c|c|c|c|}
\hline
$x_{\rm B}$ & $F_1^{\rm NC}$ & $F_2^{\rm NC}$ & $F_3^{\rm NC}$ & $F_1^{W^+}$ & $F_2^{W^+}$ & $F_3^{W^+}$ & $F_1^{W^-}$ & $F_2^{W^-}$ & $F_3^{W^-}$ \\
\hline
$ 1.0^{-5}$ & $ 6.5169^{+5}$ & $ 1.5831^{+1}$ & $ 3.9508^{+1}$ & $ 1.0443^{+6}$ & $ 1.0444^{+6}$ & $ 2.5233^{+1}$ & $ 2.5235^{+1}$ & $ 2.2026^{+4}$ & $-2.1598^{+4}$ \\
$ 1.0^{-4}$ & $ 3.0094^{+4}$ & $ 7.1530^{+0}$ & $ 1.8274^{+1}$ & $ 4.8785^{+4}$ & $ 4.8812^{+4}$ & $ 1.1524^{+1}$ & $ 1.1529^{+1}$ & $ 1.8058^{+3}$ & $-1.6079^{+3}$ \\
$ 1.0^{-3}$ & $ 1.2782^{+3}$ & $ 2.9512^{+0}$ & $ 7.9359^{+0}$ & $ 2.1053^{+3}$ & $ 2.1175^{+3}$ & $ 4.8278^{+0}$ & $ 4.8529^{+0}$ & $ 1.6680^{+2}$ & $-8.0951^{+1}$ \\
$ 1.0^{-2}$ & $ 5.0228^{+1}$ & $ 1.1089^{+0}$ & $ 3.0514^{+0}$ & $ 8.2279^{+1}$ & $ 8.7032^{+1}$ & $ 1.8084^{+0}$ & $ 1.9071^{+0}$ & $ 2.1543^{+1}$ & $ 1.1369^{+1}$ \\
$ 1.0^{-1}$ & $ 2.1078^{+0}$ & $ 4.3654^{-1}$ & $ 7.4300^{-1}$ & $ 2.6583^{+0}$ & $ 3.9864^{+0}$ & $ 5.5087^{-1}$ & $ 8.2396^{-1}$ & $ 3.0120^{+0}$ & $ 4.8797^{+0}$ \\
$ 3.0^{-1}$ & $ 3.3828^{-1}$ & $ 2.0595^{-1}$ & $ 1.6781^{-1}$ & $ 2.8339^{-1}$ & $ 6.6939^{-1}$ & $ 1.7236^{-1}$ & $ 4.0750^{-1}$ & $ 5.0084^{-1}$ & $ 1.2251^{+0}$ \\
$ 5.0^{-1}$ & $ 7.2949^{-2}$ & $ 7.3507^{-2}$ & $ 3.7503^{-2}$ & $ 4.1693^{-2}$ & $ 1.4855^{-1}$ & $ 4.1970^{-2}$ & $ 1.4969^{-1}$ & $ 8.0977^{-2}$ & $ 2.9141^{-1}$ \\
$ 7.0^{-1}$ & $ 9.7862^{-3}$ & $ 1.3749^{-2}$ & $ 4.9771^{-3}$ & $ 3.3390^{-3}$ & $ 2.0488^{-2}$ & $ 4.6882^{-3}$ & $ 2.8785^{-2}$ & $ 6.6487^{-3}$ & $ 4.0859^{-2}$ \\
$ 9.0^{-1}$ & $ 1.7778^{-4}$ & $ 3.2027^{-4}$ & $ 8.8431^{-5}$ & $ 2.0508^{-5}$ & $ 3.8298^{-4}$ & $ 3.6939^{-5}$ & $ 6.8993^{-4}$ & $ 4.1020^{-5}$ & $ 7.6590^{-4}$ \\
\hline
\end{tabular}
\end{adjustbox}\caption{NLO stucture functions with NLL evolution at $Q = 50$ GeV.}
\label{tab:N1LO-Q50}
\end{table}

\begin{table}[h]
\begin{adjustbox}{width=\textwidth}
\begin{tabular}{|c||c|c|c|c|c|c|c|c|c|}
\hline
$x_{\rm B}$ & $F_1^{\rm NC}$ & $F_2^{\rm NC}$ & $F_3^{\rm NC}$ & $F_1^{W^+}$ & $F_2^{W^+}$ & $F_3^{W^+}$ & $F_1^{W^-}$ & $F_2^{W^-}$ & $F_3^{W^-}$ \\
\hline
$ 1.0^{-5}$ & $ 9.9729^{+5}$ & $ 2.3646^{+1}$ & $ 1.0594^{+2}$ & $ 1.3337^{+6}$ & $ 1.3338^{+6}$ & $ 3.1492^{+1}$ & $ 3.1494^{+1}$ & $ 2.1969^{+4}$ & $-2.1494^{+4}$ \\
$ 1.0^{-4}$ & $ 4.3984^{+4}$ & $ 1.0220^{+1}$ & $ 4.8352^{+1}$ & $ 5.9496^{+4}$ & $ 5.9526^{+4}$ & $ 1.3756^{+1}$ & $ 1.3762^{+1}$ & $ 1.8063^{+3}$ & $-1.5895^{+3}$ \\
$ 1.0^{-3}$ & $ 1.7689^{+3}$ & $ 4.0031^{+0}$ & $ 2.0642^{+1}$ & $ 2.4329^{+3}$ & $ 2.4459^{+3}$ & $ 5.4752^{+0}$ & $ 5.5022^{+0}$ & $ 1.6843^{+2}$ & $-7.6015^{+1}$ \\
$ 1.0^{-2}$ & $ 6.4595^{+1}$ & $ 1.4056^{+0}$ & $ 7.7271^{+0}$ & $ 8.8894^{+1}$ & $ 9.3889^{+1}$ & $ 1.9269^{+0}$ & $ 2.0302^{+0}$ & $ 2.1948^{+1}$ & $ 1.2542^{+1}$ \\
$ 1.0^{-1}$ & $ 2.4151^{+0}$ & $ 4.9778^{-1}$ & $ 1.7757^{+0}$ & $ 2.6188^{+0}$ & $ 3.9463^{+0}$ & $ 5.3990^{-1}$ & $ 8.1189^{-1}$ & $ 2.9482^{+0}$ & $ 4.8573^{+0}$ \\
$ 3.0^{-1}$ & $ 3.5670^{-1}$ & $ 2.1677^{-1}$ & $ 3.7653^{-1}$ & $ 2.6060^{-1}$ & $ 6.2364^{-1}$ & $ 1.5823^{-1}$ & $ 3.7897^{-1}$ & $ 4.6037^{-1}$ & $ 1.1439^{+0}$ \\
$ 5.0^{-1}$ & $ 7.2170^{-2}$ & $ 7.2656^{-2}$ & $ 7.9652^{-2}$ & $ 3.6259^{-2}$ & $ 1.3130^{-1}$ & $ 3.6472^{-2}$ & $ 1.3220^{-1}$ & $ 7.0406^{-2}$ & $ 2.5774^{-1}$ \\
$ 7.0^{-1}$ & $ 9.0080^{-3}$ & $ 1.2651^{-2}$ & $ 9.9079^{-3}$ & $ 2.7203^{-3}$ & $ 1.6996^{-2}$ & $ 3.8183^{-3}$ & $ 2.3869^{-2}$ & $ 5.4157^{-3}$ & $ 3.3897^{-2}$ \\
$ 9.0^{-1}$ & $ 1.4497^{-4}$ & $ 2.6113^{-4}$ & $ 1.5706^{-4}$ & $ 1.4904^{-5}$ & $ 2.8369^{-4}$ & $ 2.6843^{-5}$ & $ 5.1103^{-4}$ & $ 2.9810^{-5}$ & $ 5.6734^{-4}$ \\
\hline
\end{tabular}
\end{adjustbox}\caption{NLO stucture functions with NLL evolution at $Q = 100$ GeV.}
\label{tab:N1LO-Q100}
\end{table}

\begin{table}[h]
\begin{adjustbox}{width=\textwidth}
\begin{tabular}{|c||c|c|c|c|c|c|c|c|c|}
\hline
$x_{\rm B}$ & $F_1^{\rm NC}$ & $F_2^{\rm NC}$ & $F_3^{\rm NC}$ & $F_1^{W^+}$ & $F_2^{W^+}$ & $F_3^{W^+}$ & $F_1^{W^-}$ & $F_2^{W^-}$ & $F_3^{W^-}$ \\
\hline
$ 1.0^{-5}$ & $ 5.6571^{+4}$ & $ 1.4043^{+0}$ & $ 4.3612^{-2}$ & $ 1.1411^{+5}$ & $ 1.1414^{+5}$ & $ 2.7950^{+0}$ & $ 2.7957^{+0}$ & $ 3.9818^{+4}$ & $-3.9585^{+4}$ \\
$ 1.0^{-4}$ & $ 3.3447^{+3}$ & $ 8.9927^{-1}$ & $ 1.9576^{-2}$ & $ 6.8499^{+3}$ & $ 6.8654^{+3}$ & $ 1.7990^{+0}$ & $ 1.8025^{+0}$ & $ 2.7590^{+3}$ & $-2.6549^{+3}$ \\
$ 1.0^{-3}$ & $ 2.0700^{+2}$ & $ 5.9521^{-1}$ & $ 8.8330^{-3}$ & $ 4.2120^{+2}$ & $ 4.2824^{+2}$ & $ 1.1768^{+0}$ & $ 1.1924^{+0}$ & $ 1.9262^{+2}$ & $-1.4580^{+2}$ \\
$ 1.0^{-2}$ & $ 1.5677^{+1}$ & $ 4.4505^{-1}$ & $ 3.9117^{-3}$ & $ 2.8989^{+1}$ & $ 3.2064^{+1}$ & $ 8.1638^{-1}$ & $ 8.8502^{-1}$ & $ 1.8013^{+1}$ & $ 2.6760^{+0}$ \\
$ 1.0^{-1}$ & $ 1.6533^{+0}$ & $ 3.9509^{-1}$ & $ 1.3776^{-3}$ & $ 2.2079^{+0}$ & $ 3.3478^{+0}$ & $ 5.3906^{-1}$ & $ 7.9276^{-1}$ & $ 2.9831^{+0}$ & $ 4.2298^{+0}$ \\
$ 3.0^{-1}$ & $ 4.6909^{-1}$ & $ 3.0325^{-1}$ & $ 5.0261^{-4}$ & $ 4.5052^{-1}$ & $ 9.6050^{-1}$ & $ 2.9169^{-1}$ & $ 6.2044^{-1}$ & $ 8.2373^{-1}$ & $ 1.7384^{+0}$ \\
$ 5.0^{-1}$ & $ 1.6450^{-1}$ & $ 1.7075^{-1}$ & $ 1.8147^{-4}$ & $ 1.1385^{-1}$ & $ 3.4373^{-1}$ & $ 1.1765^{-1}$ & $ 3.5687^{-1}$ & $ 2.2316^{-1}$ & $ 6.7051^{-1}$ \\
$ 7.0^{-1}$ & $ 4.0922^{-2}$ & $ 5.8220^{-2}$ & $ 4.4505^{-5}$ & $ 1.7516^{-2}$ & $ 8.7799^{-2}$ & $ 2.4837^{-2}$ & $ 1.2493^{-1}$ & $ 3.4950^{-2}$ & $ 1.7487^{-1}$ \\
$ 9.0^{-1}$ & $ 2.3255^{-3}$ & $ 4.1996^{-3}$ & $ 2.4590^{-6}$ & $ 3.4231^{-4}$ & $ 5.1469^{-3}$ & $ 6.1765^{-4}$ & $ 9.2950^{-3}$ & $ 6.8477^{-4}$ & $ 1.0292^{-2}$ \\
\hline
\end{tabular}
\end{adjustbox}\caption{NNLO stucture functions with NNLL evolution at $Q = 2$ GeV.}
\label{tab:N2LO-Q2}
\end{table}

\begin{table}[h]
\begin{adjustbox}{width=\textwidth}
\begin{tabular}{|c||c|c|c|c|c|c|c|c|c|}
\hline
$x_{\rm B}$ & $F_1^{\rm NC}$ & $F_2^{\rm NC}$ & $F_3^{\rm NC}$ & $F_1^{W^+}$ & $F_2^{W^+}$ & $F_3^{W^+}$ & $F_1^{W^-}$ & $F_2^{W^-}$ & $F_3^{W^-}$ \\
\hline
$ 1.0^{-5}$ & $ 6.7559^{+5}$ & $ 1.6167^{+1}$ & $ 5.0329^{+1}$ & $ 1.0932^{+6}$ & $ 1.0933^{+6}$ & $ 2.6001^{+1}$ & $ 2.6003^{+1}$ & $ 3.9590^{+4}$ & $-3.9043^{+4}$ \\
$ 1.0^{-4}$ & $ 3.0902^{+4}$ & $ 7.3197^{+0}$ & $ 2.0316^{+1}$ & $ 5.0548^{+4}$ & $ 5.0577^{+4}$ & $ 1.1888^{+1}$ & $ 1.1894^{+1}$ & $ 2.7462^{+3}$ & $-2.5258^{+3}$ \\
$ 1.0^{-3}$ & $ 1.2898^{+3}$ & $ 2.9968^{+0}$ & $ 8.1456^{+0}$ & $ 2.1383^{+3}$ & $ 2.1507^{+3}$ & $ 4.9301^{+0}$ & $ 4.9561^{+0}$ & $ 2.0130^{+2}$ & $-1.1319^{+2}$ \\
$ 1.0^{-2}$ & $ 4.9898^{+1}$ & $ 1.1147^{+0}$ & $ 3.0465^{+0}$ & $ 8.1772^{+1}$ & $ 8.6560^{+1}$ & $ 1.8178^{+0}$ & $ 1.9175^{+0}$ & $ 2.1548^{+1}$ & $ 1.1298^{+1}$ \\
$ 1.0^{-1}$ & $ 2.0778^{+0}$ & $ 4.3360^{-1}$ & $ 7.3305^{-1}$ & $ 2.6060^{+0}$ & $ 3.9225^{+0}$ & $ 5.4467^{-1}$ & $ 8.1639^{-1}$ & $ 2.9353^{+0}$ & $ 4.8481^{+0}$ \\
$ 3.0^{-1}$ & $ 3.3135^{-1}$ & $ 2.0256^{-1}$ & $ 1.6414^{-1}$ & $ 2.7727^{-1}$ & $ 6.5515^{-1}$ & $ 1.6938^{-1}$ & $ 4.0045^{-1}$ & $ 4.8818^{-1}$ & $ 1.1999^{+0}$ \\
$ 5.0^{-1}$ & $ 7.1573^{-2}$ & $ 7.2324^{-2}$ & $ 3.6771^{-2}$ & $ 4.1060^{-2}$ & $ 1.4567^{-1}$ & $ 4.1442^{-2}$ & $ 1.4721^{-1}$ & $ 7.9553^{-2}$ & $ 2.8569^{-1}$ \\
$ 7.0^{-1}$ & $ 9.7708^{-3}$ & $ 1.3750^{-2}$ & $ 4.9694^{-3}$ & $ 3.3652^{-3}$ & $ 2.0447^{-2}$ & $ 4.7319^{-3}$ & $ 2.8775^{-2}$ & $ 6.6926^{-3}$ & $ 4.0769^{-2}$ \\
$ 9.0^{-1}$ & $ 1.9106^{-4}$ & $ 3.4438^{-4}$ & $ 9.5055^{-5}$ & $ 2.2447^{-5}$ & $ 4.1147^{-4}$ & $ 4.0449^{-5}$ & $ 7.4166^{-4}$ & $ 4.4884^{-5}$ & $ 8.2286^{-4}$ \\
\hline
\end{tabular}
\end{adjustbox}\caption{NNLO stucture functions with NNLL evolution at $Q = 50$ GeV.}
\label{tab:N2LO-Q50}
\end{table}

\begin{table}[h]
\begin{adjustbox}{width=\textwidth}
\begin{tabular}{|c||c|c|c|c|c|c|c|c|c|}
\hline
$x_{\rm B}$ & $F_1^{\rm NC}$ & $F_2^{\rm NC}$ & $F_3^{\rm NC}$ & $F_1^{W^+}$ & $F_2^{W^+}$ & $F_3^{W^+}$ & $F_1^{W^-}$ & $F_2^{W^-}$ & $F_3^{W^-}$ \\
\hline
$ 1.0^{-5}$ & $ 1.0243^{+6}$ & $ 2.4063^{+1}$ & $ 1.3373^{+2}$ & $ 1.3811^{+6}$ & $ 1.3811^{+6}$ & $ 3.2287^{+1}$ & $ 3.2288^{+1}$ & $ 3.9344^{+4}$ & $-3.8743^{+4}$ \\
$ 1.0^{-4}$ & $ 4.5046^{+4}$ & $ 1.0453^{+1}$ & $ 5.3502^{+1}$ & $ 6.1384^{+4}$ & $ 6.1416^{+4}$ & $ 1.4165^{+1}$ & $ 1.4171^{+1}$ & $ 2.7309^{+3}$ & $-2.4909^{+3}$ \\
$ 1.0^{-3}$ & $ 1.7896^{+3}$ & $ 4.0725^{+0}$ & $ 2.1155^{+1}$ & $ 2.4738^{+3}$ & $ 2.4872^{+3}$ & $ 5.5951^{+0}$ & $ 5.6228^{+0}$ & $ 2.0185^{+2}$ & $-1.0715^{+2}$ \\
$ 1.0^{-2}$ & $ 6.4409^{+1}$ & $ 1.4149^{+0}$ & $ 7.7129^{+0}$ & $ 8.8652^{+1}$ & $ 9.3680^{+1}$ & $ 1.9392^{+0}$ & $ 2.0434^{+0}$ & $ 2.1920^{+1}$ & $ 1.2494^{+1}$ \\
$ 1.0^{-1}$ & $ 2.3847^{+0}$ & $ 4.9442^{-1}$ & $ 1.7535^{+0}$ & $ 2.5741^{+0}$ & $ 3.8905^{+0}$ & $ 5.3416^{-1}$ & $ 8.0472^{-1}$ & $ 2.8772^{+0}$ & $ 4.8285^{+0}$ \\
$ 3.0^{-1}$ & $ 3.4990^{-1}$ & $ 2.1333^{-1}$ & $ 3.6883^{-1}$ & $ 2.5537^{-1}$ & $ 6.1133^{-1}$ & $ 1.5558^{-1}$ & $ 3.7267^{-1}$ & $ 4.4933^{-1}$ & $ 1.1221^{+0}$ \\
$ 5.0^{-1}$ & $ 7.0814^{-2}$ & $ 7.1454^{-2}$ & $ 7.8100^{-2}$ & $ 3.5678^{-2}$ & $ 1.2877^{-1}$ & $ 3.5965^{-2}$ & $ 1.2994^{-1}$ & $ 6.9107^{-2}$ & $ 2.5271^{-1}$ \\
$ 7.0^{-1}$ & $ 8.9559^{-3}$ & $ 1.2595^{-2}$ & $ 9.8495^{-3}$ & $ 2.7245^{-3}$ & $ 1.6889^{-2}$ & $ 3.8289^{-3}$ & $ 2.3753^{-2}$ & $ 5.4174^{-3}$ & $ 3.3680^{-2}$ \\
$ 9.0^{-1}$ & $ 1.5308^{-4}$ & $ 2.7587^{-4}$ & $ 1.6587^{-4}$ & $ 1.5980^{-5}$ & $ 2.9948^{-4}$ & $ 2.8791^{-5}$ & $ 5.3971^{-4}$ & $ 3.1950^{-5}$ & $ 5.9890^{-4}$ \\
\hline
\end{tabular}
\end{adjustbox}\caption{NNLO stucture functions with NNLL evolution at $Q = 100$ GeV.}
\label{tab:N2LO-Q100}
\end{table}

\begin{table}[h]
\begin{adjustbox}{width=\textwidth}
\begin{tabular}{|c||c|c|c|c|c|c|c|c|c|}
\hline
$x_{\rm B}$ & $F_1^{\rm NC}$ & $F_2^{\rm NC}$ & $F_3^{\rm NC}$ & $F_1^{W^+}$ & $F_2^{W^+}$ & $F_3^{W^+}$ & $F_1^{W^-}$ & $F_2^{W^-}$ & $F_3^{W^-}$ \\
\hline
$ 1.0^{-5}$ & $ 5.6188^{+4}$ & $ 1.7533^{+0}$ & $ 4.6687^{-2}$ & $ 1.1326^{+5}$ & $ 1.1329^{+5}$ & $ 3.4231^{+0}$ & $ 3.4238^{+0}$ & $ 3.9272^{+4}$ & $-3.9021^{+4}$ \\
$ 1.0^{-4}$ & $ 3.1753^{+3}$ & $ 9.6464^{-1}$ & $ 1.9339^{-2}$ & $ 6.5344^{+3}$ & $ 6.5495^{+3}$ & $ 1.9167^{+0}$ & $ 1.9200^{+0}$ & $ 2.7201^{+3}$ & $-2.6173^{+3}$ \\
$ 1.0^{-3}$ & $ 1.9745^{+2}$ & $ 5.7885^{-1}$ & $ 8.6370^{-3}$ & $ 4.0339^{+2}$ & $ 4.1034^{+2}$ & $ 1.1475^{+0}$ & $ 1.1630^{+0}$ & $ 1.8960^{+2}$ & $-1.4389^{+2}$ \\
$ 1.0^{-2}$ & $ 1.5183^{+1}$ & $ 4.2615^{-1}$ & $ 3.8995^{-3}$ & $ 2.8075^{+1}$ & $ 3.1133^{+1}$ & $ 7.8251^{-1}$ & $ 8.5121^{-1}$ & $ 1.7838^{+1}$ & $ 2.7906^{+0}$ \\
$ 1.0^{-1}$ & $ 1.6106^{+0}$ & $ 3.9556^{-1}$ & $ 1.3675^{-3}$ & $ 2.1325^{+0}$ & $ 3.2679^{+0}$ & $ 5.3851^{-1}$ & $ 7.9410^{-1}$ & $ 2.9518^{+0}$ & $ 4.2071^{+0}$ \\
$ 3.0^{-1}$ & $ 4.5288^{-1}$ & $ 3.0005^{-1}$ & $ 4.8776^{-4}$ & $ 4.3219^{-1}$ & $ 9.2842^{-1}$ & $ 2.8837^{-1}$ & $ 6.1415^{-1}$ & $ 7.9934^{-1}$ & $ 1.6868^{+0}$ \\
$ 5.0^{-1}$ & $ 1.5979^{-1}$ & $ 1.6861^{-1}$ & $ 1.7660^{-4}$ & $ 1.1169^{-1}$ & $ 3.3374^{-1}$ & $ 1.1735^{-1}$ & $ 3.5225^{-1}$ & $ 2.1993^{-1}$ & $ 6.5102^{-1}$ \\
$ 7.0^{-1}$ & $ 4.1966^{-2}$ & $ 6.0233^{-2}$ & $ 4.5715^{-5}$ & $ 1.8563^{-2}$ & $ 8.9909^{-2}$ & $ 2.6513^{-2}$ & $ 1.2908^{-1}$ & $ 3.7085^{-2}$ & $ 1.7901^{-1}$ \\
$ 9.0^{-1}$ & $ 2.9251^{-3}$ & $ 5.2931^{-3}$ & $ 3.0959^{-6}$ & $ 4.5450^{-4}$ & $ 6.4683^{-3}$ & $ 8.2128^{-4}$ & $ 1.1705^{-2}$ & $ 9.0922^{-4}$ & $ 1.2935^{-2}$ \\
\hline
\end{tabular}
\end{adjustbox}\caption{N$^{3}$LO stucture functions with NNLL evolution at $Q = 2$ GeV.}
\label{tab:N3LO-Q2}
\end{table}

\begin{table}[h]
\begin{adjustbox}{width=\textwidth}
\begin{tabular}{|c||c|c|c|c|c|c|c|c|c|}
\hline
$x_{\rm B}$ & $F_1^{\rm NC}$ & $F_2^{\rm NC}$ & $F_3^{\rm NC}$ & $F_1^{W^+}$ & $F_2^{W^+}$ & $F_3^{W^+}$ & $F_1^{W^-}$ & $F_2^{W^-}$ & $F_3^{W^-}$ \\
\hline
$ 1.0^{-5}$ & $ 6.7295^{+5}$ & $ 1.6195^{+1}$ & $ 5.0418^{+1}$ & $ 1.0891^{+6}$ & $ 1.0891^{+6}$ & $ 2.6045^{+1}$ & $ 2.6046^{+1}$ & $ 3.9566^{+4}$ & $-3.9017^{+4}$ \\
$ 1.0^{-4}$ & $ 3.0787^{+4}$ & $ 7.3136^{+0}$ & $ 2.0289^{+1}$ & $ 5.0369^{+4}$ & $ 5.0398^{+4}$ & $ 1.1879^{+1}$ & $ 1.1885^{+1}$ & $ 2.7443^{+3}$ & $-2.5242^{+3}$ \\
$ 1.0^{-3}$ & $ 1.2857^{+3}$ & $ 2.9917^{+0}$ & $ 8.1378^{+0}$ & $ 2.1318^{+3}$ & $ 2.1442^{+3}$ & $ 4.9222^{+0}$ & $ 4.9481^{+0}$ & $ 2.0115^{+2}$ & $-1.1312^{+2}$ \\
$ 1.0^{-2}$ & $ 4.9765^{+1}$ & $ 1.1135^{+0}$ & $ 3.0471^{+0}$ & $ 8.1561^{+1}$ & $ 8.6349^{+1}$ & $ 1.8159^{+0}$ & $ 1.9157^{+0}$ & $ 2.1544^{+1}$ & $ 1.1308^{+1}$ \\
$ 1.0^{-1}$ & $ 2.0742^{+0}$ & $ 4.3359^{-1}$ & $ 7.3275^{-1}$ & $ 2.6003^{+0}$ & $ 3.9165^{+0}$ & $ 5.4456^{-1}$ & $ 8.1640^{-1}$ & $ 2.9335^{+0}$ & $ 4.8464^{+0}$ \\
$ 3.0^{-1}$ & $ 3.3064^{-1}$ & $ 2.0237^{-1}$ & $ 1.6382^{-1}$ & $ 2.7662^{-1}$ & $ 6.5380^{-1}$ & $ 1.6921^{-1}$ & $ 4.0007^{-1}$ & $ 4.8720^{-1}$ & $ 1.1976^{+0}$ \\
$ 5.0^{-1}$ & $ 7.1444^{-2}$ & $ 7.2266^{-2}$ & $ 3.6706^{-2}$ & $ 4.1023^{-2}$ & $ 1.4540^{-1}$ & $ 4.1447^{-2}$ & $ 1.4708^{-1}$ & $ 7.9485^{-2}$ & $ 2.8514^{-1}$ \\
$ 7.0^{-1}$ & $ 9.8069^{-3}$ & $ 1.3811^{-2}$ & $ 4.9883^{-3}$ & $ 3.3898^{-3}$ & $ 2.0519^{-2}$ & $ 4.7699^{-3}$ & $ 2.8899^{-2}$ & $ 6.7416^{-3}$ & $ 4.0912^{-2}$ \\
$ 9.0^{-1}$ & $ 1.9877^{-4}$ & $ 3.5839^{-4}$ & $ 9.8903^{-5}$ & $ 2.3616^{-5}$ & $ 4.2800^{-4}$ & $ 4.2567^{-5}$ & $ 7.7170^{-4}$ & $ 4.7221^{-5}$ & $ 8.5591^{-4}$ \\
\hline
\end{tabular}
\end{adjustbox}\caption{N$^{3}$LO stucture functions with NNLL evolution at $Q = 50$ GeV.}
\label{tab:N3LO-Q50}
\end{table}

\begin{table}[h]
\begin{adjustbox}{width=\textwidth}
\begin{tabular}{|c||c|c|c|c|c|c|c|c|c|}
\hline
$x_{\rm B}$ & $F_1^{\rm NC}$ & $F_2^{\rm NC}$ & $F_3^{\rm NC}$ & $F_1^{W^+}$ & $F_2^{W^+}$ & $F_3^{W^+}$ & $F_1^{W^-}$ & $F_2^{W^-}$ & $F_3^{W^-}$ \\
\hline
$ 1.0^{-5}$ & $ 1.0214^{+6}$ & $ 2.4082^{+1}$ & $ 1.3387^{+2}$ & $ 1.3773^{+6}$ & $ 1.3774^{+6}$ & $ 3.2310^{+1}$ & $ 3.2312^{+1}$ & $ 3.9326^{+4}$ & $-3.8724^{+4}$ \\
$ 1.0^{-4}$ & $ 4.4928^{+4}$ & $ 1.0446^{+1}$ & $ 5.3451^{+1}$ & $ 6.1230^{+4}$ & $ 6.1261^{+4}$ & $ 1.4155^{+1}$ & $ 1.4161^{+1}$ & $ 2.7296^{+3}$ & $-2.4898^{+3}$ \\
$ 1.0^{-3}$ & $ 1.7855^{+3}$ & $ 4.0678^{+0}$ & $ 2.1142^{+1}$ & $ 2.4685^{+3}$ & $ 2.4818^{+3}$ & $ 5.5890^{+0}$ & $ 5.6166^{+0}$ & $ 2.0175^{+2}$ & $-1.0710^{+2}$ \\
$ 1.0^{-2}$ & $ 6.4285^{+1}$ & $ 1.4139^{+0}$ & $ 7.7140^{+0}$ & $ 8.8487^{+1}$ & $ 9.3515^{+1}$ & $ 1.9379^{+0}$ & $ 2.0422^{+0}$ & $ 2.1917^{+1}$ & $ 1.2502^{+1}$ \\
$ 1.0^{-1}$ & $ 2.3816^{+0}$ & $ 4.9439^{-1}$ & $ 1.7529^{+0}$ & $ 2.5701^{+0}$ & $ 3.8862^{+0}$ & $ 5.3407^{-1}$ & $ 8.0470^{-1}$ & $ 2.8758^{+0}$ & $ 4.8272^{+0}$ \\
$ 3.0^{-1}$ & $ 3.4937^{-1}$ & $ 2.1318^{-1}$ & $ 3.6831^{-1}$ & $ 2.5495^{-1}$ & $ 6.1042^{-1}$ & $ 1.5547^{-1}$ & $ 3.7242^{-1}$ & $ 4.4868^{-1}$ & $ 1.1205^{+0}$ \\
$ 5.0^{-1}$ & $ 7.0729^{-2}$ & $ 7.1418^{-2}$ & $ 7.8006^{-2}$ & $ 3.5659^{-2}$ & $ 1.2861^{-1}$ & $ 3.5971^{-2}$ & $ 1.2987^{-1}$ & $ 6.9071^{-2}$ & $ 2.5238^{-1}$ \\
$ 7.0^{-1}$ & $ 8.9830^{-3}$ & $ 1.2640^{-2}$ & $ 9.8798^{-3}$ & $ 2.7401^{-3}$ & $ 1.6938^{-2}$ & $ 3.8528^{-3}$ & $ 2.3834^{-2}$ & $ 5.4484^{-3}$ & $ 3.3776^{-2}$ \\
$ 9.0^{-1}$ & $ 1.5791^{-4}$ & $ 2.8466^{-4}$ & $ 1.7112^{-4}$ & $ 1.6629^{-5}$ & $ 3.0888^{-4}$ & $ 2.9968^{-5}$ & $ 5.5681^{-4}$ & $ 3.3248^{-5}$ & $ 6.1771^{-4}$ \\
\hline
\end{tabular}
\end{adjustbox}\caption{N$^{3}$LO stucture functions with NNLL evolution at $Q = 100$ GeV.}
\label{tab:N3LO-Q100}
\end{table}

\clearpage
\bibliographystyle{elsarticle-num}
\bibliography{n3lo-structure-functions.bib}

\end{document}